\documentclass[twocolumn]{aastex631}

\newcommand{\lya}{Ly$\alpha$}

\newcommand{\heiL}{He\,\textsc{i}\,$\lambda$}
\newcommand{\heii}{He\,\textsc{ii}}
\newcommand{\heiiL}{He\,\textsc{ii}\,$\lambda$}
\newcommand{\ciii}{C\,\textsc{iii}]}

\newcommand{\civL}{C\,\textsc{iv}]\,$\lambda$}

\newcommand{\niiiL}{N\,\textsc{iii}]\,$\lambda$}
\newcommand{\nivL}{N\,\textsc{iv}]\,$\lambda$}

\newcommand{\ciiiL}{C\,\textsc{iii}]\,$\lambda$}

\newcommand{\civ}{C\,\textsc{iv}}
\newcommand{\oiiL}{[O\,\textsc{ii}]\,$\lambda\lambda$}
\newcommand{\oiiiL}{[O\,\textsc{iii}]\,$\lambda$}
\newcommand{\ovL}{O\,\textsc{v}]\,$\lambda\lambda$}
\newcommand{\oiiiSFL}{O\,\textsc{iii}]\,$\lambda\lambda$}
\newcommand{\neiiiL}{[Ne\,\textsc{iii}]\,$\lambda$}

\newcommand{\he}{H\,$\epsilon$}
\newcommand{\hd}{H\,$\delta$}
\newcommand{\hg}{H\,$\gamma$}
\newcommand{\hb}{H\,$\beta$}
\newcommand{\ha}{H\,$\alpha$}
\newcommand{\mgii}{Mg\,\textsc{ii}}
\newcommand{\mgiiL}{Mg\,\textsc{ii}\,$\lambda$}

\newcommand{\oiiiauL}{[O\,\textsc{iii}]\,$\lambda$}

\newcommand{\jwst}{\emph{JWST}}
\newcommand{\hst}{\emph{HST}}
\newcommand{\flux}{erg\,s$^{-1}$\,cm$^{-2}$}

\newcommand{\asec}{^{\prime\prime}}
\newcommand{\gal}{GN-$z$8-LAE}

\newcommand{\gsim}{\raisebox{-0.13cm}{~\shortstack{$>$ \\[-0.07cm] $\sim$}}~}

\usepackage{amsmath}
\usepackage{amsmath,amssymb}
\usepackage{float}
\usepackage[section]{placeins}
\usepackage{multirow}
\usepackage{tablefootnote}

\submitjournal{ApJ}

\graphicspath{{./}{}}
\shorttitle{\gal: A LyC leaker candidate at $z=8.279$}
\shortauthors{Navarro-Carrera et al.}

\begin{document}

\title{The interstellar medium conditions of a strong Ly$\alpha$ emitter at $z=8.279$ revealed by JWST: a robust LyC~leaker candidate at the Epoch of Reionization}

\correspondingauthor{Rafael Navarro-Carrera}
\affiliation{Kapteyn Astronomical Institute, University of Groningen, P.O. Box 800, 9700AV Groningen, The Netherlands}
\email{navarro@astro.rug.nl}
\author[0000-0001-6066-4624]{Rafael Navarro-Carrera}
\affiliation{Kapteyn Astronomical Institute, University of Groningen, P.O. Box 800, 9700AV Groningen, The Netherlands}d

\author[0000-0001-8183-1460]{Karina I. Caputi}
\affiliation{Kapteyn Astronomical Institute, University of Groningen, P.O. Box 800, 9700AV Groningen, The Netherlands}
\affiliation{Cosmic Dawn Center (DAWN), Copenhagen, Denmark
}

\author[0000-0001-8386-3546]{Edoardo Iani}
\affiliation{Kapteyn Astronomical Institute, University of Groningen, P.O. Box 800, 9700AV Groningen, The Netherlands}

\author[0000-0002-5104-8245]{Pierluigi Rinaldi}
\affiliation{Steward Observatory, University of Arizona, 933 North Cherry Avenue, Tucson, AZ 85721, USA}
\affiliation{Kapteyn Astronomical Institute, University of Groningen, P.O. Box 800, 9700AV Groningen, The Netherlands}

\author[0000-0002-5588-9156]{Vasily Kokorev}
\affiliation{Department of Astronomy, The University of Texas at Austin, Austin, TX 78712, USA}

\author[0000-0002-1273-2300]{Josephine Kerutt}
\affiliation{Kapteyn Astronomical Institute, University of Groningen, P.O. Box 800, 9700AV Groningen, The Netherlands}

\begin{abstract}
Making use of JWST NIRSpec and NIRCam data, we conduct a detailed analysis of \gal, a strong Ly$\alpha$ emitter at $z=8.279$. We investigate the interstellar medium (ISM) conditions that enable the Ly$\alpha$ detection in this source at the Epoch of Reionization (EoR), and scrutinize \gal as an early reionizer. In agreement with previous results, we find that \gal is a young galaxy (age $\sim 10\, \rm Myr$) with lower stellar mass ($\rm M^\star\sim 10^{7.66}\, M_\odot$) than most Ly$\alpha$ emitters at similar redshifts. The derived stellar mass and star formation rate surface densities are $\Sigma_{M^\star} \sim 355 \, \rm M_\odot/pc^2$ and $\Sigma_{\rm SFR} \sim 88\, \rm M_\odot \, yr^{-1}\, kpc^2$, respectively. Our spectral analysis indicates that: the Ly$\alpha$ line peak has a small velocity offset $\Delta v = 133 \pm 72 \, \rm km/s$ with respect to the galaxy systemic redshift; $\rm CIV] / CIII] \approx 3.3$; the ISM is characterized by a hard ionization field, although no signature of AGN is present. Moreover, we report the presence of NIII]$\lambda 1750$ emission implying super-solar N abundance, which makes \gal one of the first strong Ly$\alpha$ and nitrogen emitters at the EoR. Based on all these properties, we apply a wide range of methods to constrain the Lyman continuum escape fraction ($\rm f^{LyC}_{esc}$) of \gal and report $\rm f^{LyC}_{esc}>14\%$ in all cases. Therefore, we conclude that \gal is a robust candidate for a Lyman continuum (LyC) leaker at the EoR which is being caught at the moment of efficiently reionizing its surrounding medium.

\end{abstract}

\newcommand{\xcite}{\textcolor{red}{***}}

\section{INTRODUCTION} \label{sec:intro}

Finding the sources that reionized the Universe within the first billion years of cosmic time constitutes one of the main goals of extragalactic astronomy. Based on a number of observational constraints, it has been determined that the Epoch of Reionization (EoR) occurred between redshifts $z\approx10$ and $z\approx 6$ \citep{robertson_cosmic_2015}, although some recent quasar studies suggest that it would have only been completed by redshift $z\approx 5.3$ \citep{kulkarni_probing_2019, bosman_hydrogen_2022}. 

During this period,  the intergalactic medium changed progressively from a state dominated by neutral atomic hydrogen (HI) to a phase with virtually all atomic H ionized. This process requires that sufficient H ionizing photons (with energy $\geq 13.6 \, \rm eV$) are produced within galaxies and can escape their interstellar media to reach and ionize their circumgalactic gas. Reionization is, therefore, depicted as a process in which growing bubbles of ionized gas progressively fill the intergalactic medium until making it completely transparent \citep[e.g., ][]{dayal_early_2018}.

The opacity of the IGM during the EoR prevents direct observation of basically all the H-ionizing photons produced by galaxies at that time. Indeed, there is a declining incidence of Ly$\alpha$ emitting galaxies at $z\gsim 6$  \citep[e.g.,][]{stark_keck_2010, fontana_lack_2010, pentericci_spectroscopic_2011, caruana_spectroscopy_2014, schenker_line-emitting_2014}. However, this decline is less evident in the most luminous star-forming galaxies, with several examples of bright Ly$\alpha$  emitters found at $z\sim 7-8$ \citep[e.g.,][]{zitrin_lyman_2015, roberts-borsani_z_2016, stark_ly_2017, kumari_jades_2024, witstok_jades_2024}. These sources constitute an excellent laboratory to investigate how the patchy reionization process evolved in the first billion years of cosmic time. 

An alternative approach to understanding the drivers of reionization is to study the so-called Lyman continuum (LyC)  leakers at low and intermediate redshifts \citep[e.g.,][]{leitherer_lyman_1995, steidel_lyman-continuum_2001, vanzella_direct_2018, flury_low-redshift_2022}. These are rare star-forming galaxies in which a significant amount ofionizing photons are able to escape the source and its circumgalactic medium (CGM) to reach the observer. If the escape fraction is $f_{\rm esc}\gsim 0.1$, then the LyC leaker is considered a good analog of the sources of reionization. This is because it is estimated that star-forming galaxies at high $z$ must have had $f_{\rm esc.}\gsim 0.1$ to have efficiently conducted the reionization process \citep[e.g., ][]{rosdahl_sphinx_2018}, although some observational studies concluded that a value $\gsim 0.05$ could be enough \citep{atek_most_2024}.

As the LyC leakage cannot be directly observed at the EoR, considerable effort has been devoted to identifying other photometric and spectroscopic properties that characterize LyC leakers. These galaxies typically show compact morphologies with high star formation rate densities \citep[$\Sigma_{\rm SFR}$; e.g., ][]{naidu_rapid_2020};  a steep UV continuum; a narrow Ly$\alpha$ line with high rest-frame EW ($> 70 \, \rm \AA$)  and with a small velocity offset ($< 200 \, \rm km/s$) with respect to the galaxy systemic redshift derived from other emission lines \citep{verhamme_lyman-_2017, izotov_low-redshift_2018}; a line ratio [OIII]$\lambda 4959, 5007$/[OII]$\lambda 3727, 3729 \gsim 5$ \citep{nakajima_ionization_2014, izotov_lbt_2017, flury_low-redshift_2022}; and strong CIV emission, with C IV$\lambda1550$/C III]$\lambda1909 \gsim 0.75$ \citep{schaerer_strong_2022}, among others. However, very rarely have all these properties been reported together in LyC leakers.

With its high sensitivity up to mid-infrared wavelengths, JWST \citep{gardner_james_2023} is making the study of galaxies within the EoR routinely possible \citep[e.g.,][]{endsley_jwstnircam_2023, rinaldi_midis_2023, rinaldi_midis_2024,caputi_midis_2024, iani_midis_2024-1, bunker_jades_2023, donnan_evolution_2023, castellano_jwst_2024, harikane_jwst_2024}.  JWST spectroscopic capabilities in the infrared allow for unprecedented investigations of the interstellar media in some of these sources \citep[e.g., ][]{bunker_jades_2023, matthee_eiger_2023, schaerer_discovery_2024}. These studies point towards the presence of relatively extreme interstellar medium conditions, which are not typically found at lower redshifts \citep[e.g.,][]{mestric_exploring_2022, vanzella_extremely_2023}. With JWST spectroscopy one can search, in particular, for the signatures that characterize LyC leakers \citep[e.g.,][]{mascia_closing_2023}. Moreover, JWST spectra have enabled the discovery of unexpected emission lines whose origin is difficult to explain, such as NeIII]$\lambda1749-1753$ and  NIV]$\lambda 1486$, which has recently been reported in a number of high-$z$ objects  \citep[e.g., ][]{isobe_jwst_2023, marques-chaves_extreme_2024, schaerer_discovery_2024}.

In this work, we present a detailed spectroscopic and photometric study of \gal, a strong Ly$\alpha$ emitter at $z=8.279$ selected from the public JWST Advanced Deep Extragalactic Survey \citep[JADES;][]{eisenstein_jades_2023} in GOODS-North. This source was previously studied in other works \citep[e.g., ][]{hainline_cosmos_2024} and within a sample of high-$z$ Ly$\alpha$ emitters by \citet{witstok_jades_2024} (under the name JADES-GN-z8-0-LA) who inferred that it could have a very high escape fraction and lived in a large ionized bubble, with the likely presence of interacting companions. Here we perform an independent study of the JWST data for this source, particularly extending the analysis of spectral properties in order to obtain a more complete understanding of the physical conditions of this strong early reionizer candidate. 

Our paper is structured as follows. In Section~\ref{sec:dataset} we describe the datasets used in our study.  We present our independent derivation of the main physical properties of \gal, based on the analysis of both spectroscopic and photometric data, in Section~\ref{sec:physical properties}. In Section~\ref{sec:reioniz} we test all the main criteria to identify LyC leakers and show that \gal~complies with all of them. Finally, in Section~\ref{sec:conclusion}  we summarize our findings and present some concluding remarks. We adopt thoughout a flat $\Lambda$CDM cosmology with $\Omega_{\rm M} = 0.3$, $\Omega_\Lambda=0.7$ and $H_0 = 70 \, \rm km s^{-1} Mpc^{-1}$, and a Kroupa initial mass function \citep{kroupa_variation_2001}. In all cases, the term `escape fraction' ($\rm f_{esc.}$) refers to its absolute value.


\section{DATA} \label{sec:dataset}
\begin{figure*}[t]
\centering
\includegraphics[width=1\textwidth]{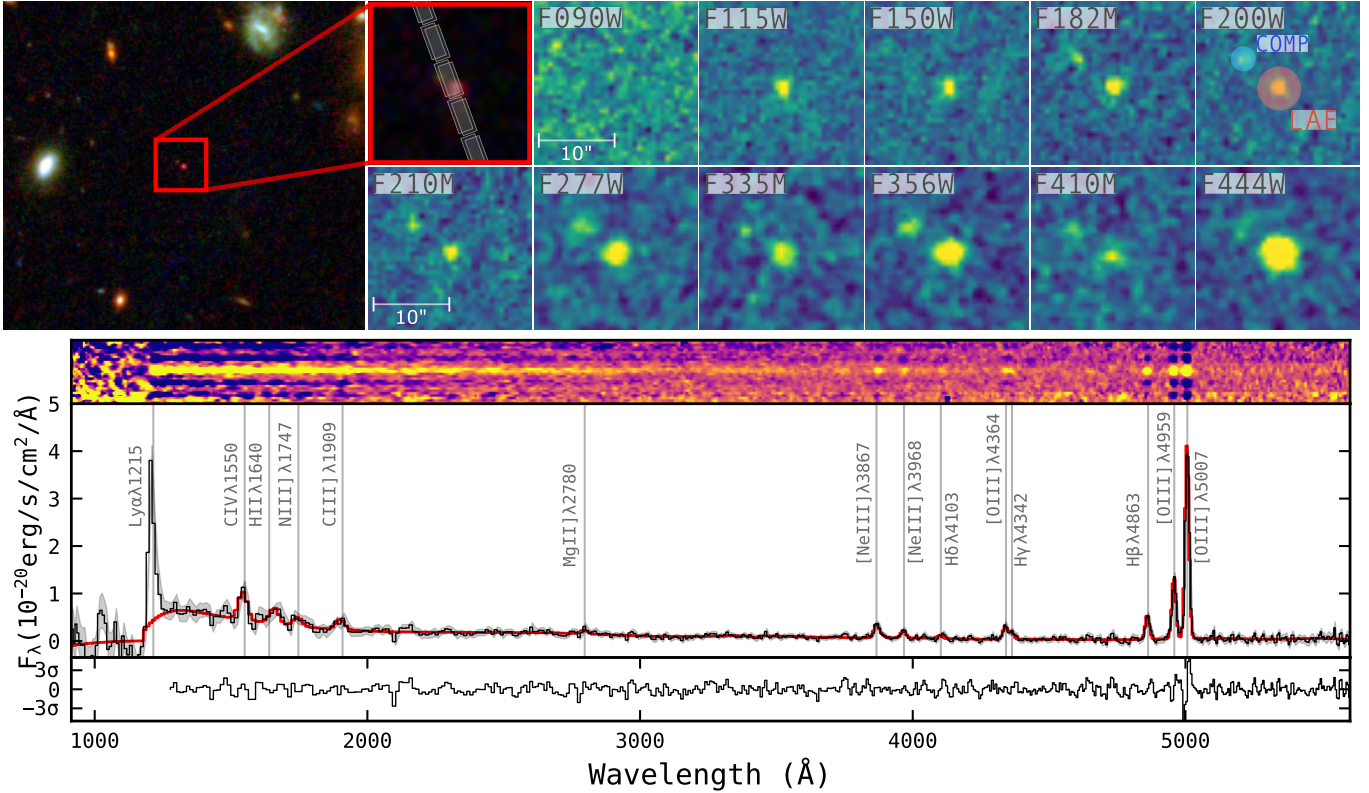}
\caption{\textbf{Top:} Postage stamps centered on \gal~for all NIRCam bands, together with the RGB image, and a zoom-in showing the position of the MSA shutters for prism observations. Note the presence of a close companion, marked as COMP. \textbf{Bottom:} 1D and 2D prism spectrum for \gal. We show the observed spectrum as a black line and the best-fit \texttt{MSAEXP} model (narrow lines and continuum) in red. Vertical lines highlight the position of some bright emission lines used in this work. We do not include the \lya\, fit, as we make use of an asymmetric profile as described in Sect. \ref{seclya}.}
\label{fig:postage}
\end{figure*}

We use the publicly available spectroscopic and photometric data from JADES \citep{eisenstein_overview_2023} in the GOODS-N, as part of the Public Data Release 3 \citep{deugenio_jades_2024}. It contains several thousands of readily available reduced, extracted, and calibrated NIRSpec/MSA spectra. We complement this rich dataset with \jwst/NIRCam and HST imaging.

In the following sections, we will discuss in more detail the observations of \gal, which is included in the JADES catalog with NIRSpec-ID 1899 and NIRCam-ID 1010260 ($\rm \alpha_{2000},\delta_{2000}$: 189h11m51.86s, 62°15'25.06").

\subsection{JWST/NIRCam and HST observations}
We collect all \jwst imaging from the JADES third release public \jwst/NIRCam data \cite{deugenio_jades_2024}. \hst \, images for this source were retrieved from the Hubble Legacy Fields (HLF) GOODS-N public data release, as described by \cite{whitaker_hubble_2019}.

We choose to extract forced photometry using Photutils \citep{bradley_astropyphotutils_2022}, using apertures of $0.\asec15$ radius to minimize possible contamination between \gal~and a nearby faint, high-$z$ source. After confirming that the position of \gal~is accurate in the JADES catalog, we use the centroids to perform forced photometry. We applied the aperture corrections derived by the JADES team for matching filters and aperture size, and corrected for galactic extinction as described in \cite{navarro-carrera_constraints_2023}.

We estimated the errors by placing a series of $0.\asec15$ radius apertures around each source in blank parts of the sky. We then measured the standard deviation between the recovered fluxes, which is the 1$\sigma$ uncertainty for our flux measurements. We use 3$\sigma$ uncertainties as upper limits and set the flux to zero in the case of non-detection.

As a sanity check, we compared our photometry against the one in the JADES catalog \citep{deugenio_jades_2024}. All our flux measurements are in agreement with the ones reported by JADES, within their respective uncertainties.

We explore the morphology of \gal~by modeling it for several NIRCam bands, covering the rest-frame UV and optical. We used both S\'ersic \citep{sersic_influence_1963}, Gaussian, and point-like PSF-convolved models, and fitted them to the NIRCam observations using \texttt{PetroFit} \citep{geda_petrofit_2022} with PSFs generated using \texttt{WebbPSF} \citep{perrin_updated_2014} in a 3x3" background subtracted cutout centered around \gal.

Our findings suggest that \gal~is only resolved in the short wavelength (SW) bands, with an effective radius of $\rm 143 \pm 15$ pc (measured in F150W, corresponding to the rest-frame UV of \gal). The fit residuals for the long wavelength (LW) bands show a good fit and do not show significant differences when using S\'ersic, Gaussian, or point-like profiles (suggesting an effective radius $\rm < 190$ pc).

A visual inspection of \gal highlights the presence of a faint source very close in projection to \gal, $3.\asec5$ in the detector plane. This galaxy is part of the JADES photometric catalog but does not have any spectroscopic observations associated with it, and could be a fainter high-$z$ candidate. Fig. \ref{fig:postage} shows the orientation of the shutters relative to\gal, together with $22\asec$ cutouts for all NIRCam bands centered in \gal~(labeled as LAE). We also highlight the companion located north-west of \gal~(labeled as COMP).

\subsection{NIRSpec/MSA observations}
We use NIRSpec/MSA spectroscopic observations from the JADES NIRCam+NIRSPec program in the GOODS-N (PID: 1181, PI: Eisenstein). The data are publicly available in the MAST (DOI: \href{https://archive.stsci.edu/doi/resolve/resolve.html?doi=10.17909/8tdj-8n28}{10.17909/8tdj-8n28})

The observations of our target were taken in two different configurations: low-resolution ($\rm R \sim 100$) using prism and covering the spectral range 0.6–5.3 $\rm \mu m$, and medium-resolution ($\rm R \sim 1000$) using the G140M, G235M, and G395M gratings. We refer the reader to \cite{deugenio_jades_2024} for a more detailed description of the observation strategy and reduction of the spectroscopic data used in this paper.

The NIRSpec observations are comprised of a total of two dithers with the 3-point nodding pattern. This translates into 12 integrations and a total of 12ks for the prism observations, and 6 integrations and a total of 6ks for each grating. This allows us to detect both strong (and faint) emission lines and continuum in the case of prism observations.

This source was not affected by any of the technical issues described in \cite{deugenio_jades_2024}, and the extracted spectrum has a good S/N, so no re-extraction was performed with a smaller number of pixels (e.g., 3 central pixels in \citealp{bunker_jades_2023, witstok_jades_2024}).

We use a modified version of \texttt{MSAEXP} \citep{brammer_msaexp_2023} line fitting algorithm to measure the line fluxes, their respective uncertainties, and observed equivalent widths for both prism and gratings. Among the additional features, our modified version of MSAEXP allows for broad emission line templates, and an enriched list of emission lines (Kokorev et al. 2024b in prep). \texttt{MSAEXP} provides a fit for all emission lines and continuum simultaneously. It makes use of single narrow and/or broad templates for emission lines and splines for the continuum. The template matrix is fitted using least squares. We determine the upper limits using the \texttt{MSAEXP} best-fit continuum and the observed flux and errors, assuming the typical line width for the rest of lines in each spectrum.

We have compared our extracted line fluxes with the ones quoted in the JADES catalog. The line fluxes are in excellent agreement for all measurements in common, well within the estimated uncertainties. These values are provided in Table \ref{tab:line_flux}. A detailed analysis of the \lya \, emission line is presented in Sect. \ref{seclya}.

\section{Physical Properties of \gal} \label{sec:physical properties}

\begin{deluxetable*}{l c c c c c r}[t]
\startdata
\centering
\tablecaption{\label{tab:line_flux} Line fluxes for \gal, expressed in units of $10^{-20}$\,\flux}
 & & \\
\textbf{Line} & $\rm Flux_{PRISM}$ & $\rm Error_{PRISM}$ & $\rm Flux_{GRATING}$ & $\rm Error_{GRATING}$ & $\rm EW_{0}$ \\ \hline
\civL1548,1550      & 228  & 36 & - & - & $52.1\pm 9.3$\\
\oiiiSFL1660,1666 + \heiiL1640$^a$ & 131 & 43 & - & - &  $37\pm 13$ \\
(\heiiL1640)$^a$    & $<$35& - & - & - &  - \\
\niiiL1749-1753     & 69   & 30 & - & - & $25\pm 11$ \\
\ciiiL1907,1909     & 69   & 28 & - & - &   $36\pm 13$\\
\mgiiL2795,2802     & 25   & 13 & - & - &   $16.3 \pm 9.1$\\
\oiiL3726,3729$^b$  &$<25$ & - & 21 & 12 &   -\\
\neiiiL3869$^c$     & -    & - & 59.8 & 9.7 &   \\
\neiiiL3967 + \he   & 35.0 & 7.0 & - & - &   $55 \pm 12$\\
\hd                 & 15.4 & 7.0 & 19.8 & 7.6 &   $28 \pm 14$\\
\hg$^d$             & 45   & 10 & 44 & 12 &   $161 \pm 27$\\
\oiiiauL4363$^d$    & 28.6 & 9.5 & - & - &   $97 \pm 21$\\
\hb                 & 93.1 & 8.0 & 76 & 12 &   $256 \pm 23$\\
\oiiiauL4959        & 230  & 10 & 232 & 17 &   $472 \pm 21$\\
\oiiiauL5007        & 715  & 14 & 749 & 23 &   $1357 \pm 26$\\
\hline
\enddata
\tablenotetext{a}{\heiiL1640 and \oiiiSFL1660,1666 are partially blended in the prism spectrum. We attempted to deblend the three line fluxes by fixing the line centers according to the spectroscopic redshift, the widths to the average one for all other (narrow) lines. However, we could only robustly set a (3$\sigma$) upper limit on the \heiiL1640, as this line is marginally separated from the \oiiiSFL1660,1666 doublet. None of the lines have coverage from any of the gratings. When using the \oiiiSFL1660,1666 flux for further calculations we consider the total \heiiL1640 + \oiiiSFL1660,1666 and the upper limit on the \heiiL1640 to derive a reasonable interval, including it as part of the error for all quantities.}
\tablenotetext{b}{Not detected in the prism spectrum, we quote the $\rm 3 \sigma$ upper limit. A hint of detection is present in the grating but with low signal-to-noise.}
\tablenotetext{c}{\neiiiL3869 is fully blended with H8 and \heiL3889 in the prism spectrum. Only \neiiiL3869 is detected in the grating.}
\tablenotetext{d}{\hg \, and \oiiiauL4363 are partially blended in the prism spectrum. However, both lines can be deblended by a careful line fit fixing the center of the lines. This is important for \oiiiauL4363, as it is not detected in the grating spectrum. Notably, the deblended and grating \hg \, line fluxes are in good agreement.}
\end{deluxetable*}

In the following sections, we will characterize the emission lines detected for \gal, its spectroscopic redshift and UV properties, metallicity, elemental abundances and several other parameters.

We report a summary of all physical properties derived for \gal~in Table \ref{sec:physical properties}.

\subsection{Emission lines and spectroscopic redshift} \label{sec:lines}

The combination of broad spectral coverage and high signal-to-noise from the prism spectrum allows us to detect a wealth of rest-frame optical and UV emission lines (see Table \ref{tab:line_flux}). The continuum is also well-detected at all wavelengths. Fig. \ref{fig:postage} shows both 2D and 1D spectra, the aspect of \gal~in \jwst/NIRCam images, and the positioning of the NIRSPec/MSA shutters.

We report line flux measurements and rest-frame equivalent widths (henceforth abbreviated $\rm EW_0$) in Table \ref{tab:line_flux}. 
When available, we report the measured fluxes from both prism and gratings. Notably, $\rm EW_0$s are always measured using the PRISM spectrum, as the continuum is not present in gratings. 

Interestingly, blue-wards of Ly$\alpha$ we tentatively detect an emission feature that falls at the expected wavelength of the line complex \ion{O}{6}$\lambda\lambda1032,1038$ + Ly$\beta$ + \ion{C}{2}$\lambda\lambda1036,1037$. However, a careful analysis of the spectral data reduction and extraction should be performed to confirm the presence of these lines more robustly.

The low spectral resolution of the prism observations ($\rm R \sim 100$) makes several lines appear partially or completely blended. This is shown in Table \ref{tab:line_flux}. Crucially, the blending between \hg \, and \oiiiauL4363 is only partial, and the respective line fluxes were recovered by a careful fit in which the line central wavelengths were fixed. As a sanity check, the comparison between the deblended \hg \, flux, and the one from the grating spectrum shows an excellent agreement.

The spectroscopic redshift of \gal~was derived by averaging the positions of all emission lines detected in the prism spectrum (excluding blended lines and the Ly$\alpha$ line). We report a redshift $z_{spec} = 8.2790 \pm 0.0005$. This spectroscopic redshift solution derived using NIRSpec/MSA is in good agreement with the one obtained by the FRESCO team using only the \oiiiL5007,4959 lines \citep[$z_{spec}=$8.28][]{oesch_jwst_2023}) and with the one reported by \citet{witstok_jades_2024}.

Wavelength calibration offsets between prism and grating spectra were reported by \cite{deugenio_jades_2024}, thus we double-checked the spectroscopic redshift obtained from grating G395M, with the most detected emission lines. We retrieved a $z_{spec} = 8.279 \pm 0.002$, compatible with the PRISM solution. The peak of the \oiiiauL5007 from the G395M is in excellent agreement with the redshift determined from the prism spectrum (see Fig. \ref{fig:lya}). Accordingly, no systematic effect on wavelength calibration was found between gratings and prism for \gal.

\begin{deluxetable}{l|lr}[t]
\startdata
\tabcolsep=2.75mm
\tablecaption{\label{tab:sedfit_params} Quantities derived using the spectro-photometric \texttt{BAGPIPES} SED fit, and their respective priors.}
 & & \\
\textbf{Parameter} & Reference/Prior  & Best-fit Value\\
\hline
Templates & \texttt{BC03}$^a$  & \\ 
SFH & delayed exponential  & \\ 
$e-$folding time ($\tau$, Gyr) & 0.0001–15  & $0.06\pm0.06$  \\
\hline
Mass ($\log_{10}\mathcal{M}/\mathcal{M}_{\odot}$) & $1 - 13$  & $7.66\pm 0.02$\\
Metallicity ($\rm Z/Z_{\odot}$) & $0.001 - 1$  & $0.2\pm 0.01$ \\
Age (Gyr) & 0.0001 – 15  & $0.001\pm 0.001$ \\
$\rm \log U$ & $-4 - -0.5$   & $-1.0 \pm 0.12 $  \\
\hline
Extinction law & SMC$^c$  &    \\
Av & 0.0 - 7.0  & $0.03\pm 0.02$  \\
$\beta$-slope & - & $-2.5\pm 0.1$ \\
\hline
IMF & \citet{kroupa_variation_2001}  & \\
\enddata
\tablenotetext{a}{2016 version of \citet{bruzual_stellar_2003}}
\tablenotetext{c}{\citet{gordon_quantitative_2003}}
\end{deluxetable}

\subsection{Spectro-photometric SED fitting} \label{sec:sedfit}

To fully explore the high-quality photometric and spectroscopic data available for \gal, we use \texttt{BAGPIPES} \citep{carnall_how_2019} for performing a joint spectro-photometric SED fitting. Table \ref{tab:sedfit_params} lists the configuration parameters employed for the run. Briefly, \texttt{BAGPIPES} uses stellar emission from \cite{bruzual_stellar_2003} with an \cite{kroupa_variation_2001} initial mass function (IMF) with a cut-off mass of $\rm 120 \ M_\odot$, and nebular emission from \texttt{CLOUDY} \citep{ferland_2013_2013}. We used broad and flat (uninformative) priors for all parameters, as the combination of spectrum and photometry has enough constraints without any additional information. All runs were performed by fixing the redshift to the spectroscopic value retrieved as described in the previous sections.

We use the techniques described by \cite{carnall_vandels_2019} to take full advantage of the combination of spectroscopic and photometric datasets. In particular, we allow for a $<2\sigma$ perturbation to the spectrum in the form of a second-order Chebyshev polynomial, which can correct for systematic issues in the flux calibration, although we found the applied correction to be small in the case of \gal. We also allow for a multiplicative factor on the spectroscopic errors to correct for underestimated uncertainties. These new parameters are fitted simultaneously to the rest of the physical parameters.

The best-fit results are shown in Table \ref{tab:sedfit_params}, assuming a delayed exponential star formation history \citep[see][]{schaerer_discovery_2024}. They indicate that \gal~has a low stellar mass ($\rm \mathcal{M} \sim 10^{7.66}$), is extremely young (best-fit age of $\sim 10$ Myr), and has low dust extinction ($\rm E(B-V) \sim 0.02$). The metallicity we retrieve is approximately 1/5th solar ($\rm 12 + \log(O/H) \sim 8$). Finally, the ionization parameter is elevated, with $\log U \sim -1$.

These parameters appear to be in broad agreement with the ones derived from direct spectroscopic measurements for \gal~(Table \ref{tab:properties}) and other high-$z$ LAEs (Table \ref{tab:lit_properties}) and low-$z$ LAEs \citep{ouchi_observations_2020}.

\subsection{UV magnitude and slope from the NIRSpec spectrum} \label{secuvmag_slope}

We measure an UV-magnitude of $\rm M_{UV} = -19.55 \pm 0.22$  at $\rm 1500 \,\AA$ and UV $\beta$-slope $\beta = -2.48 \pm 0.23$ \cite[using the best-fit continuum and the definition by][]{calzetti_dust_1994}. This places \gal~close to the typical $M_{UV}^\star$ at $z\sim 8$ \citep[e.g.,][]{bouwens_new_2021, donnan_evolution_2023}, making \gal~relatively bright compared to similar LAEs \citep[e.g.,][]{saxena_jades_2023}. Additionally, we retrieve the values for $\rm M_{UV}$ and UV $\beta$ slope from NIRCam+\hst \, photometry. We recover $\rm M_{UV} = -19.59 \pm 0.15$ and $\rm \beta = -2.73 \pm 0.35$. The agreement in $\rm M_{UV}$ ensures that there is no significant systematic offset between photometry and spectrum \citep[e.g., due uncertainties in path-loss corrections][]{kokorev_uncover_2023}. The departure in the $\beta$ values can be explained by the photometry being affected by strong UV emission lines.

When comparing to the median value of high-$z$ LAEs and SFGs samples studied using \jwst\, \citep[Table \ref{tab:lit_properties},][]{kumari_jades_2024,morishita_enhanced_2024,iani_midis_2024} or the general population of $z>6$ galaxies \citep{bouwens_uv_2009}, \gal~shows a bluer UV $\beta$-slope. Several studies link blue $\beta$ slopes (and absence of dust) to a enhanced $\rm f_{esc}^{LyC}$ \citep[e.g.,][]{chisholm_far-ultraviolet_2022}.

As a sanity check, the values for spectroscopic redshift, $M_{UV}$ and $\beta$ slope derived in our analysis are in good agreement with the ones reported by \citep{witstok_jades_2024}. Our $\beta$ slope is slightly bluer but in agreement within the errors (this discrepancy could be rooted in the different techniques we used to fit the FUV continuum). Our values for emission line fluxes are in good agreement with the ones reported in the JADES catalog \citep{deugenio_jades_2024}. The measurements reported by \citet{witstok_jades_2024} for \gal are systematically smaller when compared to the ones from JADEs and our study. This difference could be explained by the different extraction procedures used by \citet{witstok_jades_2024}.

\begin{deluxetable}{l r r}[t]
\tabcolsep=2.5mm
\startdata
\tablecaption{Summary of the physical properties of \gal~\label{tab:properties}}\\
Quantity & Value & Section \\
\hline
$z$                                         & $8.2790 \pm 0.0007$ & §\ref{sec:lines}\\
$\rm M_{UV}$ (mag)                          & $-19.6 \pm 0.3    $ & §\ref{secuvmag_slope}\\
$\beta$                                     & $-2.48  \pm 0.23  $ & "\\
$\rm E(B-V)$ (mag)                          & $0.02  \pm 0.38   $ & §\ref{sec:basic_props}\\
$\rm \log_{10}(M^\star/M_\odot)^\dagger$            & $7.66  \pm 0.02   $ & §\ref{sec:sedfit}\\
\hline
$\rm R_{eff}^{UV}$ (pc)                     &  $143\pm 15$        &  §\ref{sec:dataset}  \\
$\rm SFR_{H\alpha} \ (M_\odot)$             &  $11.3 \pm 7.8$     &  §\ref{sec:basic_props}   \\
$\rm SFR_{UV}\ (M_\odot)$                   &  $3.0 \pm 2.1$      &   "  \\
$\rm age\ (Myr)$                            &  $2-6$      &   "  \\
$\rm \Sigma_{M^\star}\ (M_\odot/pc^2)^\dagger$      &  $355 \pm 75$      &   §\ref{sec:n-emitters}  \\
$\rm \Sigma_{SFR(H\beta)}\ (M_\odot/yr/kpc^2)$  &  $88 \pm 32$        &   §\ref{sec:n-emitters}   \\
\hline    
$\rm T_e$ (K)                               &  $(1.73\pm 0.24)\times 10^4$  &  §\ref{sec:temp_metal}   \\
$\rm 12+ \log(O/H)_{mean}^\ddagger$                  &  $7.85 \pm 0.17$              &  "   \\
$\rm \log(C/O)$                             &  $-0.69 \pm 0.21$ &  §\ref{sec:c_abundance}   \\
$\rm \log(N/O)$                             &  $-0.44\pm0.36$   &  §\ref{sec:n-emitters}  \\
\hline
$\rm F_{Ly\alpha} \ $($10^{-20}$\,\flux)$^{\star}$       &  $753 \pm 56$   &   §\ref{seclya}  \\
$\rm EW_0^{Ly\alpha} \ (\AA)^{\star \star}$         &  $235 \pm 10$   &   "  \\
$\rm \Delta v_{Ly\alpha}$ (km/s)            &  $132 \pm 72$   &   "  \\
$\rm f_{esc}^{Ly\alpha}$                    &  $0.25-0.40$    &   " \\
\hline  
$\rm f_{esc}^{LyC}$$^{\star \star \star}$   &  $\sim 60 \% (14 \%)$  & §\ref{sec:fesc}  \\
$\rm \log \xi_{ion}^0$                      &  $25.78 \pm 0.31$       &  §\ref{sec:xion}   \\
\enddata
\tablenotetext{\dagger}{Stellar mass derived using spectro-photometric \texttt{BAGPIPES} SED fit, see Tab. \ref{tab:sedfit_params}.}
\tablenotetext{\ddagger}{Average value between direct ($\rm T_e$ based method) and optical metallicity calibrations, see Tab. \ref{tab:metal}.}
\tablenotetext{\star}{From the best-fit asymmetric model \citep{shibuya_what_2014} flux for the G140M spectrum.}
\tablenotetext{\star \star}{\, Derived using the prism spectrum, see \ref{seclya}.}
\tablenotetext{\star\star\star}{\, \, Average value for all estimators used in Sect. \ref{sec:fesc}, in brackets the smallest value.}
\end{deluxetable}

\subsection{Basic physical properties from the NIRSpec spectrum} \label{sec:basic_props}

We estimate the color excess $\rm E(B-V)$ by inspecting the Balmer decrement between \hg\, and \hb \citep[using theoretical values from][]{storey_recombination_1995}. We use an SMC \citep{gordon_quantitative_2003} extinction law for all our $\rm E(B-V)$ measurements. Accordingly, we recover a color excess of $\rm E(B-V) = 0.02 \pm 0.38$. Together with the insights from the $\beta$ slope \citep{meurer_dust_1999} and \texttt{BAGPIPES}, all three estimators converge towards the low dust content of \gal. This is in line with the results for strong \lya \, emitters in the literature \citep[e.g.,][]{rosani_bright_2020, saxena_jades_2023}. For all the reasons above, we do not apply any dust correction to line ratios, fluxes, or derived quantities in this paper.

\begin{figure}[t]
\centering
\includegraphics[width=1\columnwidth]{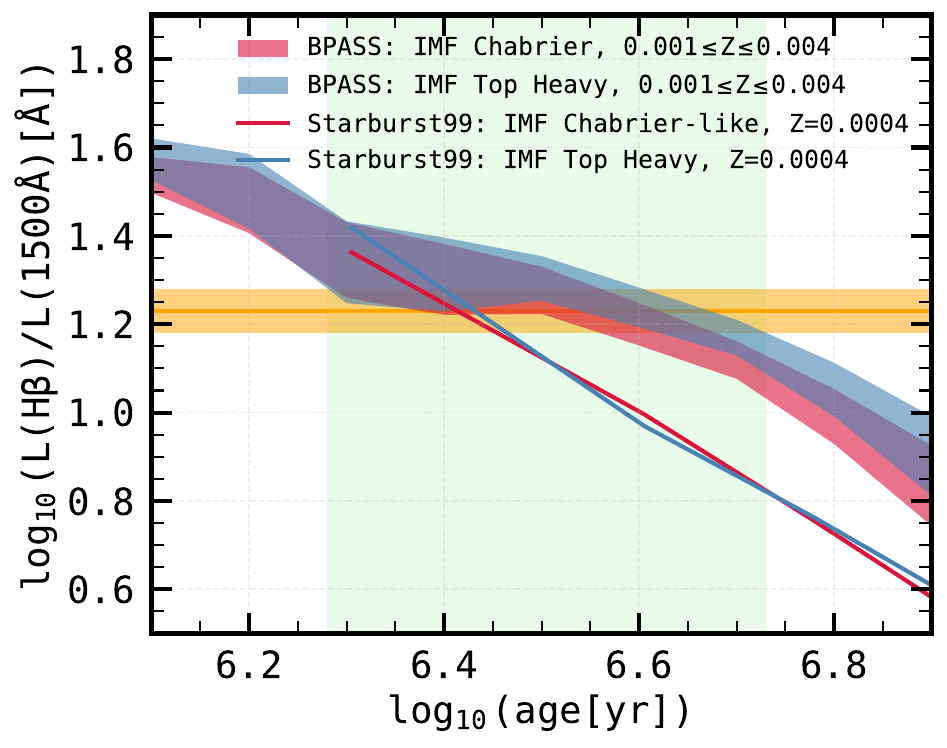}
\caption{\label{fig:age} Constraint on the light-weighted age of our target source based on the \hb \, luminosity and FUV($1500 \, \AA$) luminosity density ratio. We include theoretical tracks for a single burst SFH (since the onset of star formation) for BPASS \citep[v2.2.1][as shaded areas, between $\rm 0.001 < Z < 0.004$]{eldridge_binary_2017} and \texttt{Starburst99} \citep[][as continuous lines for a fixed $\rm Z = 0.0004$]{leitherer_starburst99_1999}. These tracks are derived for Chabrier-like and Top-heavy IMF with a cut-off mass of $\rm 300 \, M_\odot$, color-coded in red and blue, respectively (see Sect. \ref{sec:n-emitters}). The age range compatible with the line luminosity ratio measured for \gal \, (orange line) is shown as a green-shaded area.}
\end{figure}

Under the assumption that all emission lines originate from the reprocessing of ionizing photons from young, short-lived massive stars in the ISM of the galaxy (see Sect. \ref{sec:ionization}), it is possible to estimate the instantaneous star formation rate (in the past $10$ Myr) from the luminosity of Balmer lines \citep{kennicutt_star_1998,rinaldi_midis_2023} after taking into account the  \cite{kroupa_variation_2001} IMF we use in this work. 

As the PRISM spectrum does not cover \ha \, we derive the SFR from the \hb \, flux instead, by applying the relations by \cite{storey_recombination_1995}. By doing so, we retrieve $\rm L_{H\alpha} = (2.14\pm 0.18)\times 10^{42} \ erg/s$, which in turn corresponds to $\rm SFR_{H\alpha} = 11.3\pm7.8 \ M_\odot/yr$, where the error incorporates in quadrature the intrinsic scatter inherent to the \cite{kennicutt_star_1998} calibration.

If deriving the SFR from FUV luminosity density \citep{kennicutt_star_1998}, the ratio $\rm SFR_{H\alpha}/SFR_{UV}$ takes the value $\rm \log_{10} SFR_{H\alpha}/SFR_{UV} = 0.57 \pm 0.42$. As discussed by \cite{weisz_modeling_2012,faisst_recent_2019,emami_closer_2019} and Navarro-Carrera et al. (in prep.), the ratio $\rm SFR_{H\alpha}/SFR_{UV}$ is a tracer of recent burstiness in the star formation history. Ratios $\rm \log_{10} SFR_{H\alpha}/SFR_{UV} \gsim 0.2$ indicate a very recent burst in star formation, produced not longer than $50$ Myr ago. 

\gal shows a $\rm SFR_{H\alpha}/SFR_{UV}$ well above unity. This is in agreement with predictions from simulations \citep{sparre_starbursts_2017,ceverino_firstlight_2018} and observations (Navarro-Carrera et al., in prep.), where low-mass galaxies are expected to show increasingly bursty SFHs and, thus, populate the starburst cloud \citep{rinaldi_emergence_2024}. The expected light-weighted age for \gal is extremely young: $\leq 50$ Myr.

The specific star formation rate (sSFR) of \gal~is  $\rm \log_{10} sSFR_{H\alpha}/yr^{-1} = -6.55 \pm 0.30$. Following the criterion from \cite{caputi_star_2017,caputi_alma_2021}, \gal~is classified as a starburst (SB) galaxy, as $\rm \log_{10} sSFR/yr^{-1} >> -7.6$.

By further exploring the relation between the Balmer lines and the UV luminosity density it is possible to estimate a value for the light-weighted age, under the assumption of \gal~being a star-forming galaxy. We explore the evolution of stellar tracks from \texttt{Starbust99} \citep{leitherer_starburst99_1999} and \texttt{BPASS} \citep[v2.2.1][]{eldridge_binary_2017, stanway_re-evaluating_2018} for Chabrier-like and top-heavy IMFs (the uncertainty on our measurements is much larger than the one produced by the choice of IMF). We then explore the time-evolution of the ratio between $\rm L_{H\beta}$ and $\rm L^\nu_{1500}$. We refer the reader to \cite{iani_scrutiny_2022} for an in-depth discussion of the methodology. For \gal~we found a light-weighted age in the range of $2-6$ Myr, as shown in Fig. \ref{fig:age}.

All properties are in line with the ones obtained in the analysis by \citet{witstok_jades_2024}. We infer lower dust content and slightly higher instantaneous SFR (both compatible within the uncertainty).

Having reviewed the basic properties of \gal, in the following sections, we will overview in more detail our derivations for gas-phase metallicity, ionization source, and C and N abundances.
 
\vspace{10 mm}

\subsection{Auroral line determination of electron temperature, and gas-phase metallicity} \label{sec:temp_metal}

\begin{deluxetable}{l|cc}[t]
\tabcolsep=2.75mm
\startdata
\tablecaption{\label{tab:metal} Values of electron temperature and gas-phase metallicity from the direct method, UV, and optical calibrations.}
& \\
\textbf{Calibration} & $\rm 12+\log O/H$ &Reference\\
\hline
Direct & $(7.77-7.79) \pm 0.14$ & I06$^a$ \\
$T_e = (17 \pm 0.2)\times 10^3 K$&   &   \\
\hline
C3O3 & $(7.12-7.20)\pm 0.27$ & M22$^b$\\
EWCIII & $7.20 \pm 0.24$ & M22$^b$\\
\hline
R23 & $(7.86-8.03)\pm 0.10$ & S24$^c$\\
O3 & $7.92 \pm 0.10$ & S24$^c$\\
\hline
\enddata
\tablenotetext{a}{I06, \cite{izotov_chemical_2006}.}
\tablenotetext{b}{M22, \citep{mingozzi_classy_2022}.}
\tablenotetext{c}{S23, \citep{sanders_direct_2024}.}
\end{deluxetable}

By following \cite{proxauf_upgrading_2014}, we constrain the $\rm T_e(OIII)$ (hereafter $T_e$) of the ionized gas thanks to the presence of the auroral \oiiiauL4363 line (ruled by collisional excitation). In particular, we use the ratio (\oiiiauL4958,5007)/\oiiiauL4363, which is monotonically correlated to $T_e$. For \gal, we report $T_e = (1.73 \pm 0.24)\times 10^4 K$.

We use a variety of UV and optical calibrations for determining the gas-phase metallicity of \gal, together with the estimation from the direct ($T_e$) method \citep{sanders_direct_2024,izotov_chemical_2006}. For the so-called direct method, we consider that $O/H = O^+/H^+ +O^{++}/H^+$ by assuming the rest of higher ionization terms to be negligible \citep{sanders_direct_2024}. We use the calibrations from \cite{izotov_chemical_2006} to derive $O^+/H^+$ and $O^{++}/H^+$. The term $O^+/H^+$ directly depends on the \oiiL3727/\hb \, ratio, for which we have only an upper limit. We report the derived metallicity as an interval.

All the values for metallicities are displayed in Table \ref{tab:metal}. We find an average value of $\rm 12 + \log O/H = 7.85 \pm 0.17$ or $\rm Z = (0.14 \pm 0.05) Z_\odot$ \citep{asplund_chemical_2021} as an average between the two optical diagnostics and the direct method. The discrepancy between optical/direct metallicities and UV metallicities could be due to the strong dependence of UV lines on the ionization of the ISM \citep{maiolino_re_2019}.

\begin{figure*}[ht]
\centering
\plottwo{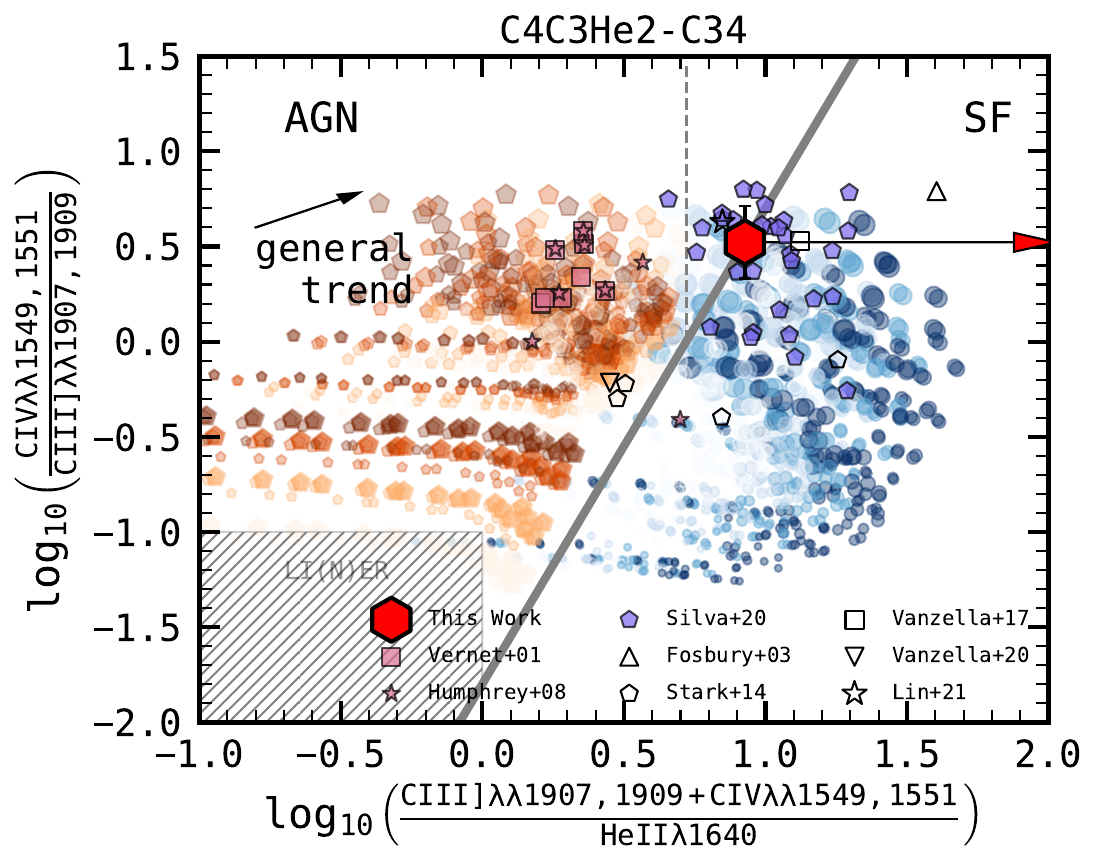}{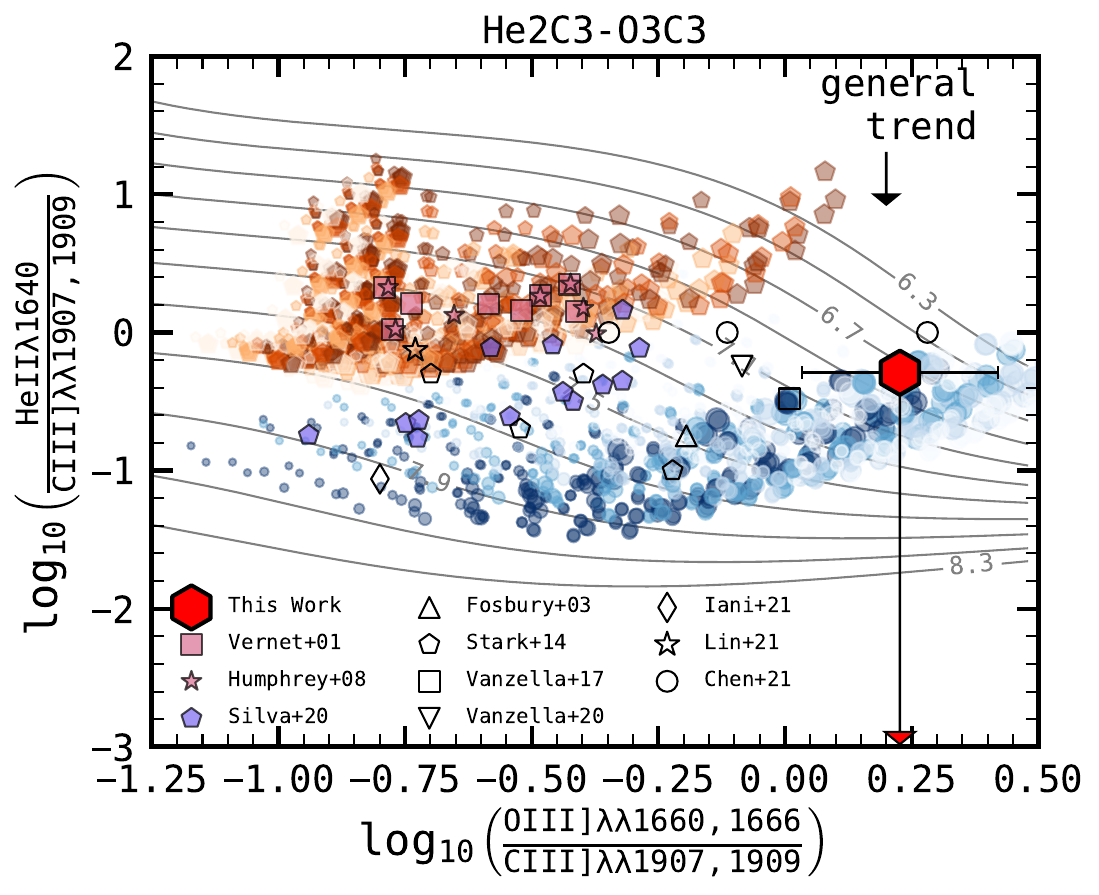}
\caption{\label{fig:agn} \textbf{Left/Right:} The C4C3He2-C34/He2C3-O3C3 line diagnostics \citep{nakajima_vimos_2018} that discriminate between AGN and SF-dominated ionized regions. The values for \gal~are shown as a red hexagon (with the upper limit on the \heii). Together with this, we include the AGN and SFG by \cite{feltre_nuclear_2016} (F16) and \cite{gutkin_modelling_2016} (G16) as red and blue markers, respectively. We show bigger markers for higher values of  $\rm \log U$, within the range $\rm \log U -4 \sim -0.75$. We refer the reader to \cite{iani_scrutiny_2022} for a description of the literature in the figures. $\rm \log(\rm C/O)$ abundance or UV spectral slope ($\alpha$, $\rm f_\nu \propto \nu^\alpha$) are related to the opacity of the markers, higher opacity was assigned to bigger values in both cases. We do not apply additional cuts to the F16 models shown in this work (either in their gas density or dust-to-metal mass ratios)}
\end{figure*}

\begin{deluxetable*}{l c c c c c c c c c c}[ht]
\tabcolsep=0.5mm
\startdata
\tablecaption{Properties of some of the high-$z$ galaxies studied with NIRSPec, including SFGs, LAEs, and N-emitter galaxies. Indirect (e.g., best-fit derived quantities) or photometric measurements are shown between brackets. \label{tab:lit_properties}}\\
Galaxy & $z$ & $\rm M_{UV}$ & $\rm \log M^\star$ & SFR & $\beta$ & $\rm EW^0_{Ly\alpha}$ & $\rm 12+\log(O/H)$ & $\rm R_{eff}$ & type & reference \\
  &   & (mag) & ($\rm M_\odot$) & ($\rm M_\odot/yr$)  &   & ($\rm \AA$) &   & (pc) & &\\
\hline
\gal~& 8.279 & -19.6 & 7.66 &  11.3 & -2.73 & 203 & 7.85 & 140 & LAE \& N-em & This Work \\
\hline
GN-z11   & 10.6 & -21.5 & 8.73  & 24   & -2.4 & 18 & 7.66     &  $64$ & LAE \& N-em & \cite{bunker_jades_2023}\\
JADES-GS-z7-LA  & 7.28 & -17.0 & 6.9  & 0.3-0.6   & -2.0 & 388 & (8.16)     &  - & LAE & \cite{saxena_jades_2023}\\
COLA1  & 6.59 & -21.35 & 9.93  & 10.1   & (-3.2) & - & 7.88     &  $<260$ & LAE & \cite{torralba-torregrosa_anatomy_2024}\\
JADES LAEs  & $>6.3$ & -19.00 & -  & -   & -2.34 & - & 7.50   &  - & LAEs & \cite{kumari_jades_2024}\\
\hline
JADES SFGs  & $>6.3$ & -19.08 & -  & -   & -2.26 & - & 7.73   &  - & SFGs & \cite{kumari_jades_2024}\\
NIRSPec SFGs  & $>5$ & -19.25 & 8.48  & 3.86   & -2.23 & - & -   &  309 & SFGs & \cite{morishita_enhanced_2024}\\
GS-z14-0 & 14.3 & -20.81 & 8.7  & 22  & -2.20 & - & (7.11) & 260 & SFG & \cite{carniani_shining_2024}\\
GS-z14-1 & 13.9 & -19.00 & 8.0  & 2   & -2.71 & - & (7.69) &  $<160$ & SFG & \cite{carniani_shining_2024}\\
GHz2     & 13.9 & -20.53 & 9.05 & 5.2 & -2.46 & - & 7.26   &  105 & SFG & \cite{castellano_jwst_2024}\\
\hline
GN-z9p4  & 9.4 & -21.00 & 8.7  & 64   & - & - & 7.37         &  $118$ & N-em & \cite{schaerer_discovery_2024}\\
RXCJ2248  & 8.05 & -20 & 8.05  & 63 & -2.72 & - & 7.43     &  $<22$ & N-em & \cite{topping_metal-poor_2024}\\
\enddata
\end{deluxetable*}

By inspecting Table \ref{tab:lit_properties}, \gal~is slightly more metal-rich when compared to typical LAEs or N-emitters at similar redshifts \citep{kumari_jades_2024,schaerer_discovery_2024,topping_metal-poor_2024}. However, we find it consistent with the SFGs population \citep{kumari_jades_2024} and with the expectations from the mass-metallicity relation \citep[MZR, although slightly above the median trend reported by][]{curti_jades_2024}.

\subsection{The ionization source} \label{sec:ionization}

The presence of UV emission lines is usually associated with hard ionization fields and ionization parameters in the range $\rm \log U \geq -2.5$ \footnote{$\rm \log U$ is defined as the ratio between the number density of ionizing photons and the number density of hydrogen atoms \citep[e.g.,][]{osterbrock_astrophysics_2006}}, coupled with relatively low gas-phase metallicity ($\rm 12+\log (O/H) \leq 8.3$) \citep{jaskot_photoionization_2016,mingozzi_classy_2024}. The presence of strong high-ionization lines such as \civL1548,1550 in the spectrum of \gal \, (highly likely from nebular origin) strengthens this idea.

The \oiiiauL5007/\oiiL3727 ratios probe intermediate-ionization regions. By using UV diagnostics calibrated for these regions \citep{mingozzi_classy_2022}, we get  $\rm \log U = -1.36 \pm 0.5$  from $\rm EW_0$(\civL1549), and $\rm \log U = -1.35 \pm 0.32$ from  \civL1548/\ciiiL1907. These values are compatible with the one derived in the spectro-photometric SED fit (Sect. \ref{sec:sedfit}). When compared to the median values of $\rm \log U$ of high-z SFGs \citep[$z\sim 1.5-6$][]{reddy_jwstnirspec_2023,reddy_impact_2023}, \gal~shows an enhanced $\rm \log U$, more in line with the results of \cite{williams_magnified_2023} for an ultra-compact, magnified galaxy at $z=9.51$.

We discard the presence of a Type I Active Galactic Nucleus (AGN) \citep{mccarthy_high_1993,ramos_almeida_testing_2011}, as no evidence of broadening or wings has been found for \gal. This is true for both the prism and grating spectra, with essentially all lines being unresolved ($\rm \Delta V \leq 200$ km/s).  The average measured line width for the G395M/F290LP spectrum (the one with more detected lines) is $\sim 210 \pm 175$ km/s. Moreover, no X-ray detection in the vicinity of \gal~is reported within the 2 Ms Chandra Deep Field-North Survey catalog \citep{xue_2_2016}. However, we discuss the possibility of a Type II AGN in the following paragraphs.

Several UV diagnostics have been proposed to discriminate between SF or AGN-dominated galaxies. In Fig. \ref{fig:agn} we show two diagnostics adapted from \cite{iani_first_2022}—namely C4C3-C34 and C3He2-O3C3 \citep[e.g.][]{nakajima_vimos_2018, hirschmann_synthetic_2019, byler_comparison_2020}. We overlay photoionization models for nebular emission in SFGs \citep[as blue hexagons,][G16]{gutkin_modelling_2016} and AGN \citep[red pentagons,][F16]{feltre_nuclear_2016}. We limit the models to subsolar metallicities ($\rm Z = 0.0005-0.006 = (0.025-0.3) \, Z_\odot$).

\gal~falls in the SF region for C4C3He2–C34 and He2C3-O3C3 (see Fig. \ref{fig:agn}), in particular, the one with the highest ionization parameter. However, it does it only marginally in the case of C4C3He2–C4C3 and also when putting it in the context of the mass-excitation diagram \citep[MEX,][]{juneau_active_2014}. We also find \gal~to be compatible with SF galaxies with the highest $\rm logU$ by using the  O3\hg/O33 diagnostic \citep{mazzolari_new_2024}. 

In light of these results, the ionizing radiation in \gal~probably originated from a population of young, massive stars, rather than AGN. However, a contribution of AGN, shocks, and SF \citep[so-called composite][]{hirschmann_synthetic_2019} can not be completely ruled out. 


\subsection{C/O abundance} \label{sec:c_abundance}

Following \cite{perez-montero_using_2017}, we estimate the C/O abundance from the ratio between (\ion{C}{3}]$\lambda\lambda 1907,1909$ + \ion{C}{4}$\lambda1548$) and \ion{O}{3}]$\lambda\lambda 1660,1666$) \citep[see also][]{berg_carbon_2016}. We estimate a $\log(\rm C/O) = -0.69 \pm 0.21$, i.e. almost half of the solar value \citep[{[C/O]}$_\odot = 0.44$,][]{gutkin_modelling_2016}. 

\begin{figure}[t]
\centering
\includegraphics[width=1\columnwidth]{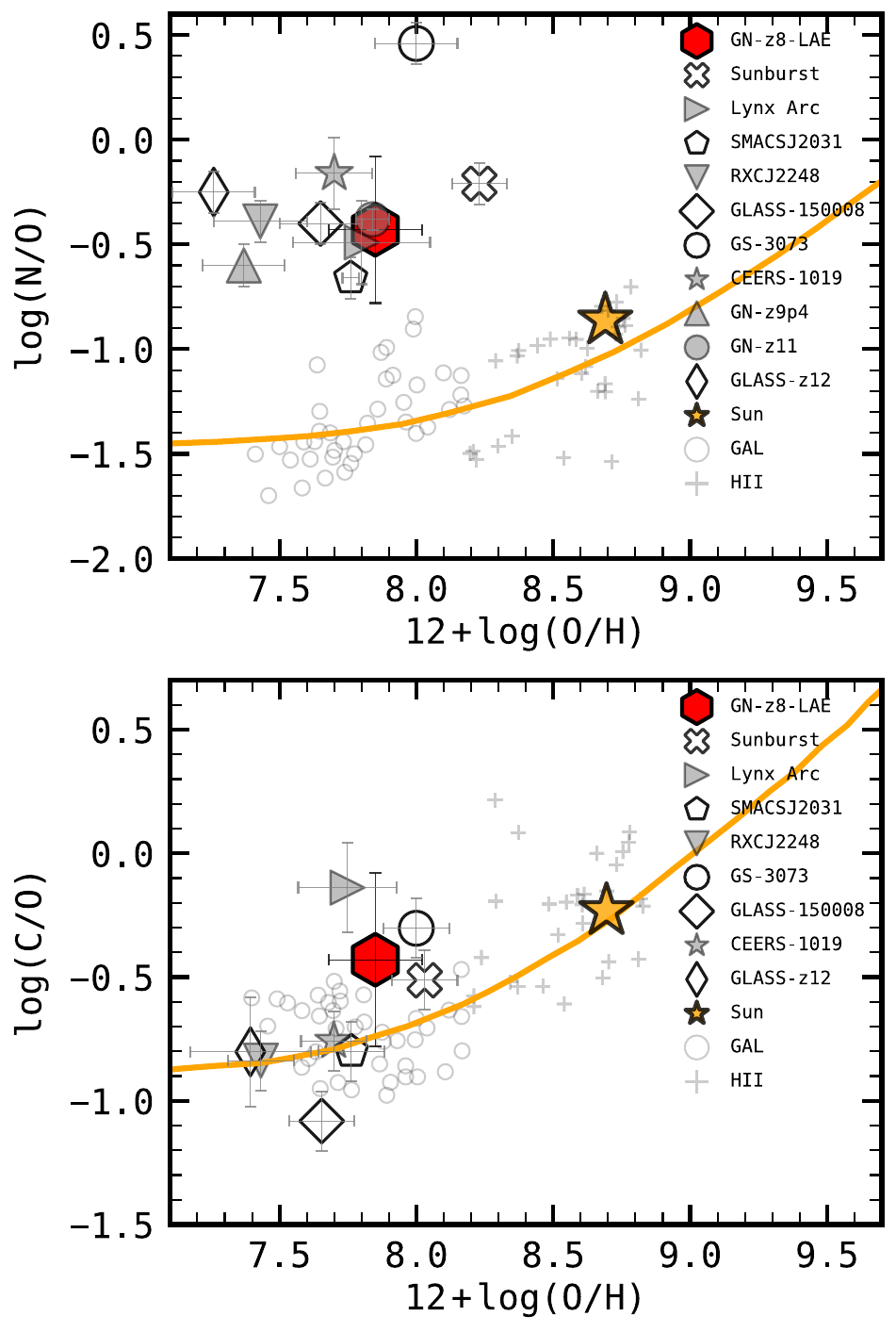}
\caption{ \textbf{Top/bottom:} Nitrogen/carbon over oxygen abundance as a function of metallicity, respectively. We include points for Lynx and Sunburst arc \citep{villar-martin_nebular_2004,mestric_exploring_2022}, all high-$z$ N-emitters proven by NIRSpec \citep{marques-chaves_extreme_2024,topping_metal-poor_2024,ji_ga-nifs_2024,isobe_jwst_2023,castellano_jwst_2024}, and set of low-redshift HII regions and galaxies from \cite{izotov_lyman_2022}. The expected trend from local studies \citep{vila-costas_nitrogen--oxygen_1993,dopita_modeling_2006} is shown as an orange line, and the solar values \citep{asplund_chemical_2021} are included as an orange star.}
\label{fig:n_emitters}
\end{figure}

The value we retrieved is in agreement with those of local, metal-poor dwarf galaxies \citep{berg_chemical_2019} and with young low-mass and low-luminosity galaxies at intermediate redshift ($z = 1.5 - 5.0$) \citep[e.g.,][]{erb_physical_2010,stark_ultraviolet_2014}. Our target metallicity estimate (12 + $\log_{10}(\rm O/H) \sim 7.85$) is in agreement with the (C/O) - metallicity trend reported by \citep{garnett_evolution_1995, nicholls_abundance_2017}.  

The physical origin of this trend is likely associated with the metallicity dependence of winds from massive rotating stars, along with the delayed release of carbon from lower mass stars \citep[e.g.,][]{henry_cosmic_2000, akerman_evolution_2004}.

\subsection{The \niiiL1749-1753 \, complex \label{sec:n-emitters}}
Several high-$z$ N-emitter galaxies have been found very recently using JWST/NIRSpec: GHZ2/GLASS-z12 \citep{castellano_jwst_2024}, GN-z11 \citep{bunker_jades_2024, cameron_jades_2023, senchyna_gn-z11_2024},  GN-z9p4 \citep{schaerer_discovery_2024}, GLASS-150008 \citep{isobe_jwst_2023}, CEERS-1019 \citep{marques-chaves_extreme_2024}, RXCJ2248-4431 \citep{topping_metal-poor_2024} and GS-3073 \citep{ji_ga-nifs_2024}.

As reported in Table \ref{tab:line_flux}, we find a \niiiL1749-1753 (blended complex of emission lines) $2.3 \sigma$ detection in the PRISM spectrum of \gal. This would position \gal~amongst the highest redshift N-emitter galaxies, and the only one with strong \lya\, emission. When inspecting the PRISM spectrum of \gal~we found a weak feature resembling an emission line in the expected position of \nivL1486, however, it is not possible to constrain it further due to its low signal-to-noise.

\begin{figure}[t]
\centering
\includegraphics[width=1\columnwidth]{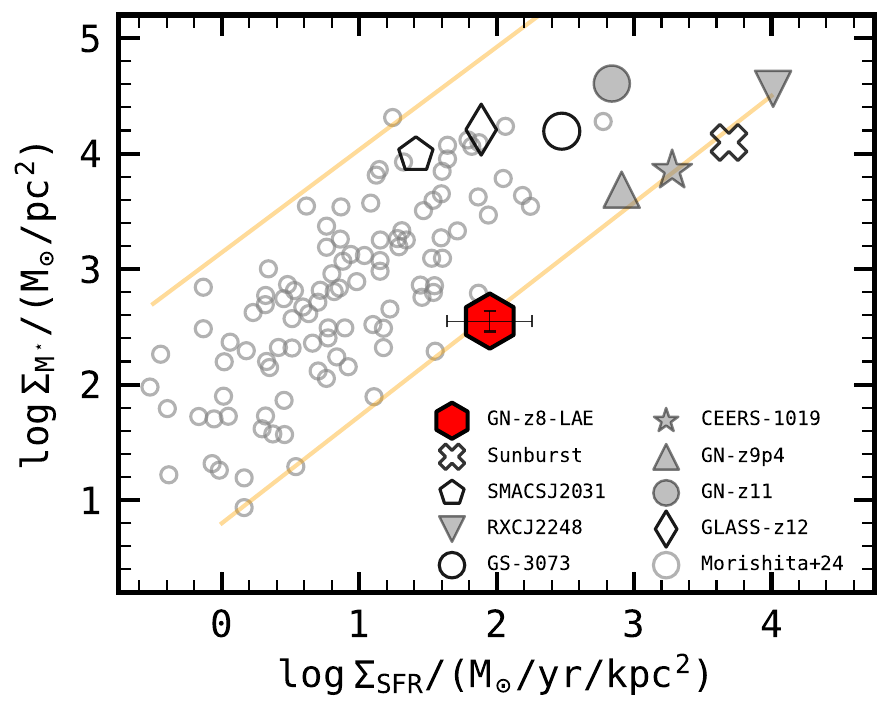}
\caption{\label{fig:sfr_density} Stellar mass surface density as a function of star formation rate surface density. Here, \gal~is shown as a red hexagon. We include the Sunburst arc \citep{mestric_exploring_2022} and all high-$z$ N-emitters proven by NIRSpec \citep{marques-chaves_extreme_2024,topping_metal-poor_2024,ji_ga-nifs_2024,isobe_jwst_2023,castellano_jwst_2024}. The orange lines mark the sSFR envelope for SFGs and N-emitters, and portrays \gal~with the maximum sSFR together with some of the N-emitters, that have higher stellar mass.}
\end{figure}

By using the relations from \citep{villar-martin_nebular_2004,aller_physics_1984}, we derived $\rm \log (N^{2+}/H+) = -4.6\pm0.3$ directly from our measurement of \niiiL1749-1753/\hb. As we do not have any detection of higher ionization N lines (\nivL1486), we use the relations by \cite{hamann_metallicities_2002} for the metallicity of \gal~to predict the \nivL1486 flux from our \civL1548,1550 measurement. This allows us to determine $\rm \log (N^{3+}/H+) = -6.4\pm0.4$. The total Nitrogen abundance is dominated by the contribution of the direct measurement of $\rm N^{2+}/H+$, so the assumptions behind the determination of $\rm N^{3+}/H+$ could only further increase the Nitrogen abundance.

All in all, we derive an elevated, super-solar Nitrogen abundance of $\rm \log(N/H) = -4.58\pm0.36$ and, in turn, $\rm \log(N/O) = -0.44\pm0.36$ for \gal. In particular, this ratio translates to approximately 2.8 times the solar ratio \citep{asplund_chemical_2021}. This is in line with other high-$z$ N-emitters compact SFGs, characterized by an N-enriched (with supersolar N abundances) and metal-poor (low-metallicity, O/H) ISM \citep{marques-chaves_extreme_2024}.

As of today, no unique scenario exists to interpret the presence of N lines and reconcile the super-solar nitrogen abundance in metal-poor galaxies.  Wolf-Rayet stars, very massive and supermassive stars ($\rm M^\star > 100 \, M_\odot$), a top-heavy IMF, and tidal disruption events have been suggested as mechanisms that could enhance the nitrogen abundance \citep[see][and references therein]{schaerer_discovery_2024}.

In the case of \gal, the strong high-ionization lines, \lya \, and \niiiL1749-1753, seem to favor the scenario of a short burst of intense star formation in the present or past few million years \citep{topping_metal-poor_2024}. This is in agreement with our findings in Sec. \ref{sec:sedfit} and \ref{sec:physical properties}.

Taking advantage of the morphological properties we derived in \ref{sec:dataset}, we can study the position of \gal~in the stellar mass surface density ($\rm \Sigma_M^\star$) - star formation surface density ($\rm \Sigma_{SFR}$) plane, shown in Fig. \ref{fig:sfr_density}. We report these results in Table \ref{tab:properties}, following \cite{reddy_jwstnirspec_2023} and using the SFR derived from the Balmer lines and the stellar mass from the spectro-photometric fit reported in Table \ref{tab:sedfit_params}.

In Fig. \ref{fig:sfr_density} we include the compilation of spectroscopically confirmed SFGs at $z>5$ by \citep{morishita_enhanced_2024}, and all the known Nitrogen emitters discovered with NIRSpec (see Sect. \ref{sec:n-emitters}). From the position in this plane, \gal~shows a smaller stellar mass surface density compared to the rest of N-emitters, and is located in the extreme sSFR envelope of the SFG distribution (again, as several other N-emitters).
\subsection{The \lya \, emission of JADES-GN-z8-LA} \label{seclya}

\gal~was selected as a strong \lya\, emitter, as can be seen in Fig. \ref{fig:postage}. We detect \lya \, in both prism and the G140M grating spectra with a signal-to-noise of approximately 12 in both cases.

We model the \lya \, line using an asymmetric profile as proposed by  \cite{shibuya_what_2014}. We report a flux of $\rm F_{Ly\alpha} = (753\pm56)\times 10^{-20} \, erg/s/cm^2$ for the G140M grating. We also report a $\rm EW_0 = 235 \pm 10 \, \AA$. Note that this value was derived using the prism spectrum, where the spectral continuum is detected.

\begin{figure}[t]
\centering
\includegraphics[width=1\columnwidth]{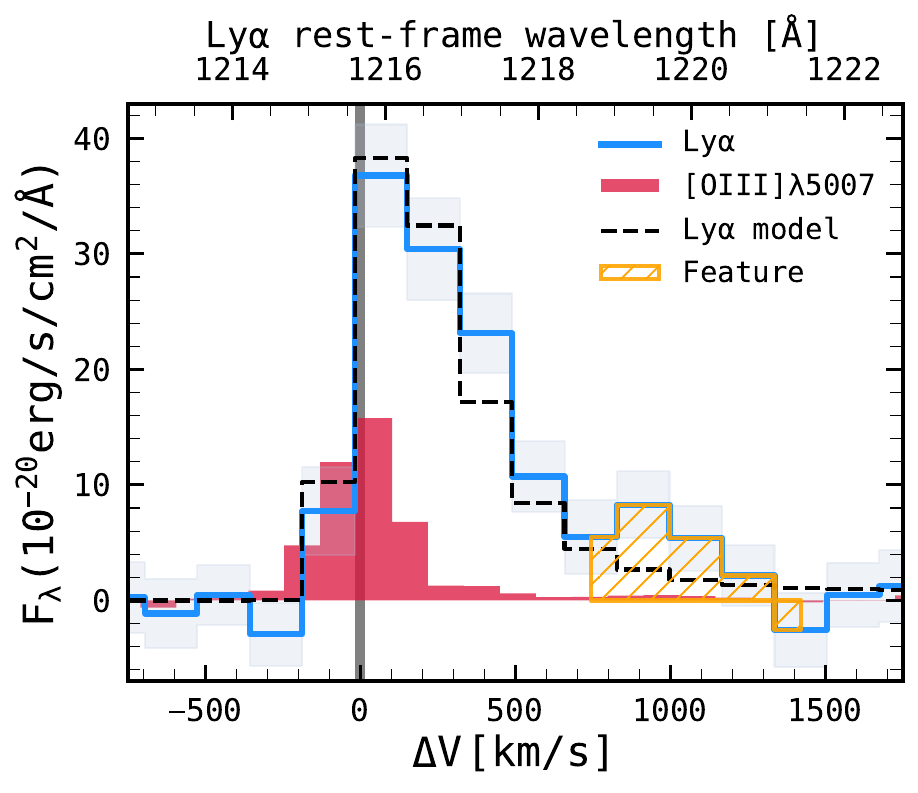}
\caption{\label{fig:lya} The \lya ~line from the grating (G140M) spectrum in velocity space (blue line). The asymmetric model from \cite{shibuya_what_2014} is shown as a dashed black line. Relative to it, the \oiiiSFL5007 line profile from the G395M grating observations is shown as a red area. The orange hatched area highlights the emission feature investigated in Sect. \ref{seclya}.}
\end{figure}

From the best-fit \lya ~model, we report a velocity offset of $\rm \Delta v_{Ly\alpha} = 133 \pm 72$ km/s with respect to the systemic velocity of the system. This value is small, within the line spread function of NIRSpec around $1 \, \rm \mu m$ \citep{saxena_jades_2023}, but similar to the one reported by \cite{saxena_jades_2024} for several LAEs and also compatible with values from the compilation of $z>5.8$ LAEs by \cite{witstok_inside_2024}. The presence of \lya \, emission close (or at) the systemic velocity could indicate a low neutral hydrogen column density over the line-of-sight \citep[LOS][]{kerutt_lyman_2024}.

The \lya \, line is not resolved in the prism spectrum, and we find a similar situation for the G140M grating (Fig. \ref{fig:lya}). We recover a best-fit line width comparable to the spread function (LSF), of $\rm \sim 200 \, km/s$.

The flux recovered from the prism spectrum is $\rm F_{Ly\alpha} = (871\pm82)\times 10^{-20} \, erg/s/cm^2$. This value is enhanced with respect to the model flux from the G140M grating. We explain this by the presence of a feature red-wise of the \lya \, line. This feature is blended with \lya \, for the prism observations, but can be resolved using the G140M grating, as shown in Fig. \ref{fig:lya} (orange hatched area). We found several possible explanations for this feature assuming it is not a random fluctuation: 1. A broad red wing associated with the \lya \, line \citep{yang_z_2014}. 2. An effect of the \textsc{OV]}1218 emission line \citep{humphrey_o_2019,iani_first_2022}. 3. An effect of \lya \, scattering.

Several arguments disfavor the possibility of the feature being \textsc{OV]}1218 emission. First of all, we find no evidence of the presence of the remaining component of the doublet \ovL1214,1218. \textsc{OV]}1214 is expected to be brighter than \textsc{OV]}1218 from models \citep{humphrey_o_2019} and it is far enough from \lya \, to be resolved in the G140M spectrum. Perhaps more importantly, we found the line center associated with \textsc{OV]}1218 not to be in agreement with the systemic redshift of \gal.

The specific shape of the \lya \, profile is expected to depend on the conditions of both intergalactic medium (IGM) and ISM \citep{blaizot_simulating_2023, witten_evidence_2023, witten_deciphering_2024}. However, a detailed modeling of the \lya \, profile falls out of the scope of our analysis.

By comparing the ratio between the observed \lya \, flux, and Balmer lines to theoretical predictions, we can estimate the \lya \, escape fraction \citep[although with caveats, see][]{choustikov_physics_2024}. We assume a Case B recombination and an electron density between 100 $\rm cm^{-3}$ and 1000 $\rm cm^{-3}$ and use the electron temperature derived in Sect. \ref{sec:temp_metal}. Using \texttt{PyNeb} \citep{luridiana_pyneb_2015}, we derive  $\rm F_{Ly\alpha}/F_{H\beta} = 25.68 - 31.80$, for 100 $\rm cm^{-3}$ and 1000 $\rm cm^{-3}$, respectively. By comparing this ratio with the one observed in \gal~we retrieve a $f^{esc}_{Ly\alpha} = 0.25-0.40$. Any dust correction would shift $f^{esc}_{Ly\alpha}$ towards higher values.

Our derived $f^{esc}_{Ly\alpha}$ is in agreement with the one found by using indirect tracers \citep[e.g., $\rm EW_0^{Ly\alpha}$, $\beta$ and $\xi_{ion}$][]{sobral_predicting_2019}. In particular, we employ the relation corrected for the ionizing photon production efficiency (see Sect. \ref{sec:xion}) and dust extinction from \citet[][ Eq.~6]{sobral_predicting_2019} to retrieve $\rm f^{esc}_{Ly\alpha} = 0.63 \pm 0.31$. This value is compatible with the direct method estimations within the error bars.

Our retrieved value for the \lya \, escape fraction is reasonable when comparing it to a sample of LAEs at $5.8<z<7.98$ selected from JADES NIRSPec observations \citep{witstok_inside_2024}. In particular, an enhanced $f^{esc}_{Ly\alpha}$ could also be linked to an enhanced $f^{esc}_{LyC}$  \citep{chisholm_accurately_2018, naidu_synchrony_2022}.

\section{\gal~as a L\MakeLowercase{y}C leaker} \label{sec:reioniz}

Young and star-forming galaxies such as \gal~are expected to produce a high number of ionizing photons, but only the ones that escape the galaxy (i.e., are not absorbed in ISM) effectively contribute towards the reionization of the universe \citep{chisholm_optically_2020}. In particular, LyC leakers are defined by having a non-negligible escape of LyC photons.

Two scenarios for the escape of LyC photons have been proposed, even though low-mass/faint and star-forming galaxies are rich in neutral hydrogen, which is opaque to LyC radiation \citep[e.g.,][]{watts_relationship_2021}. The density-bounded medium scenario involves the presence of a low-column-density neutral hydrogen medium surrounding the sources of ionizing radiation. Some authors have suggested that it fails to reproduce spectral properties of LyC leakers \citep{ramambason_reconciling_2020}.

The second scenario consists of the presence of an optically thick but porous medium, that does not completely cover all the sources of ionizing radiation \citep{zackrisson_spectral_2013}. Following \cite{rivera-thorsen_sunburst_2017}, we divide this second scenario into two cases: one where the medium is clumpy and provides a multitude of clear LOS \citep[picket fence model,][]{gronke_lyman-alpha_2016}, another where there are only a few channels open for the LyC photons to escape \citep[ionized channels model,][]{zackrisson_spectral_2013}. In principle, the \lya \, profile (shaped by the resonant interactions with the neutral hydrogen), could show evidence of these two different geometries \citep[][and references therein]{rivera-thorsen_sunburst_2017, hutter_astraeus_2023}.

For a given galaxy, the number of ionizing (Lyman continuum, LyC) photons injected into the IGM per unit time can be expressed as
\begin{equation}
    \rm \dot{N}_{ion} = f_{esc}^{LyC}\xi_{ion}L_{UV} \,,
\end{equation}
Where $\rm f^{LyC}_{esc}$ (escape fraction of ionizing photons) represents the fraction of LyC photons that escape the galaxy with respect to the total amount produced by the young stellar population (in the case of \gal). In turn, $\xi_{ion}$ (hydrogen ionizing photon production) can be understood as the number of hydrogen ionizing photons per unit of UV luminosity \citep[e.g.,][]{naidu_synchrony_2022, lin_empirical_2024}, and is defined as:
\begin{equation}
    \rm {\xi_{ion} = \frac{L_{H\beta}}{4.76\times 10^{-13} (1-f_{esc}) L^{int}_{UV}}} \,.
\end{equation}

From the above relations, it can be seen that $\rm f_{esc}^{LyC}$ is a key parameter to determine $\rm \dot{N}_{ion}$, and $\rm \xi_{ion}$. However, the LyC emission can rarely be observed directly for $z>4$ galaxies, as the mean-free path for LyC photons decreases exponentially with redshift \cite[][]{inoue_updated_2014}.

In the following sections, we will explore the possible contribution of galaxies like \gal~to the reionization of the Universe and, particularly, the possibility of \gal~efficiently injecting ionizing photons into the IGM.

\subsection{Tracers of high LyC escape fraction} \label{sec:fesc_tracers}

The literature has extensively explored the estimation of $\rm f_{esc}^{LyC}$ using different tracers and techniques. Most studies rely on small samples of spectroscopically confirmed LyC leakers low-$z$ analogs or simulations to predict whether a galaxy is a LyC leaker based on one or several of its observed properties.

\begin{figure}[t]
\centering
\includegraphics[width=1\columnwidth]{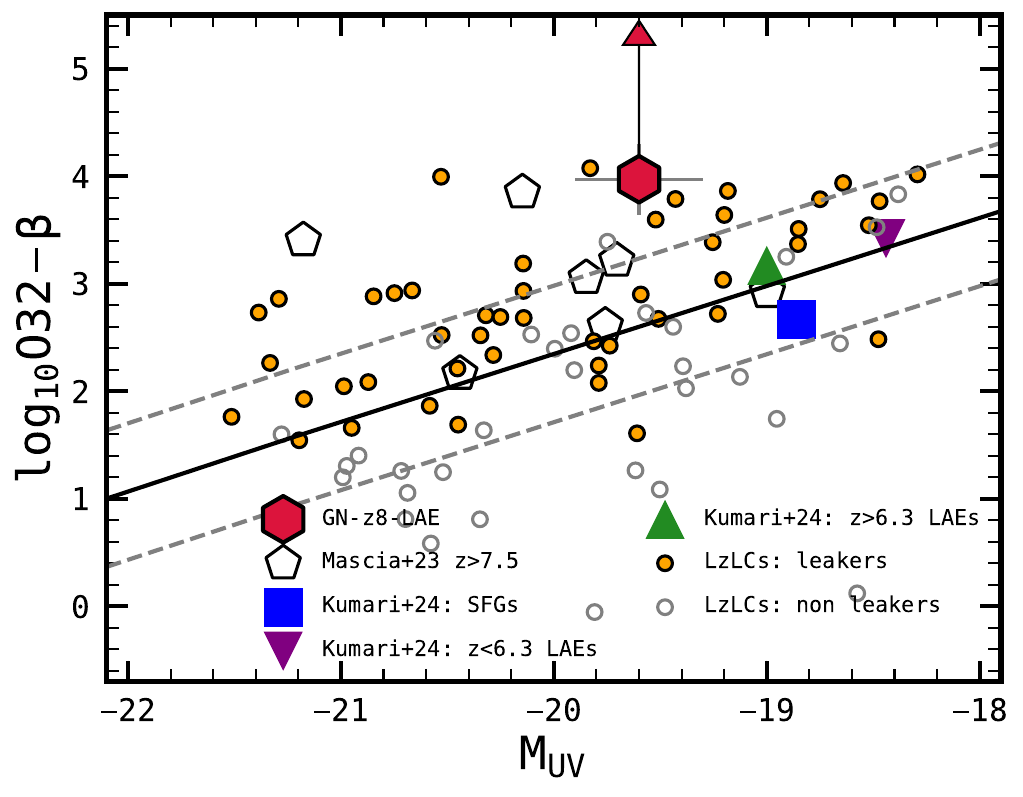}
\caption{\label{fig:lin_fesc} Diagnostic for classifying LyC leakers based on $\rm \log_{10}O32 - \beta$ vs $\rm M_{UV}$ as presented in \citet{lin_empirical_2024}. The continuous line indicates the $50 \%$ probability of leaking LyC, while the discontinuous lines indicate the $25 \%$ and $75 \%$ values. \gal~is shown as a red hegaxon. We include the LzLCs divided into leakers and not leakers, the $z>7.5$ simple compiled by \cite{mascia_closing_2023}, and the average values for JADES SFGs, $z<6.3$ LAEs and $z>6.3$ LAEs reported in \cite{kumari_jades_2024}.}
\end{figure}

Among the most relevant single-variable indicators for elevated LyC leakage, we find:
\begin{itemize}
\item Properties of the \lya \, emission: \cite{verhamme_lyman-_2017} found that LyC leakers preferently have \lya\, large rest-frame equivalent widths ($\rm EW_0^{Ly\alpha}>70 \, \AA$) and escape fractions ($\rm f_{esc}^{Ly\alpha} > 20 \%$). \\ More recent studies \citep[e.g.,][using low-$z$ analogs]{naidu_synchrony_2022,izotov_lyman_2021} suggest that leakers are characterized by a high $\rm f_{esc}^{Ly\alpha}$, and more crucially, narrow \lya \, profiles close to the systemic velocity (low values of $\rm \Delta v_{Ly\alpha}$). This would indicate a highly transparent IGM. \citet{chisholm_accurately_2018} provides a estimation for $\rm f_{esc}^{LyC}$ based on the dust-corrected $\rm f_{esc}^{Ly\alpha}$.

\item Dust content or $\beta$ slope: Several works suggest that LyC leakers are characterized by blue $\beta$ slopes and low dust content \citep[][for samples of low-z analogs]{chisholm_far-ultraviolet_2022, saldana-lopez_low-redshift_2022}. In particular, $\beta$ was found to correlate with several parameters apart from dust content, such as \oiiiauL5007/\oiiL3726,9 (O32) ratio, but also metallicity, $\rm EW_0^{H\beta}$, or stellar mass. 

\item $\rm \Sigma_{SFR}$: Using the X-SHOOTER Lyman-$\alpha$ survey  \citep[XLS-z2,][]{matthee_x-shooter_2021} sample, \cite{naidu_rapid_2020}  argue that intense and compact star formation (high $\rm \Sigma_{SFR}$) might create the necessary conditions (ionized channels) that allow for an efficient escape of LyC (and \lya) photons \citep[see][$\rm f_{esc}^{LyC}$ would be a feedback-regulated process]{pucha_ly_2022, trebitsch_fluctuating_2017}.

\item Elevated \civ/\ciii \, ratio and $\rm EW_0^{CIV}$: The presence of strong nebular \civ  \, (from its $\rm EW_0$ or the \civ/\ciii \, ratio) has been identified as a proxy for enhanced $\rm f_{esc}^{LyC}$
 \citep[e.g.,][for samples of $z\sim3-4$ SFGs and $z\sim0.3-0.4$ LyC leakers]{saxena_strong_2022, schaerer_strong_2022}.

\item Elevated \oiiiSFL5007/\oiiL3726,3729 (O32) ratio: \cite{izotov_lbt_2017, flury_low-redshift_2022} found evidence of a correlation between the
O32 line ratio for samples of low-$z$ analogs \citep[although this ratio depends strongly on the ionization parameter,][]{strom_measuring_2018}. More recent studies, however, seem to suggest O32 is not necessarily elevated for LyC leakers, or that the correlation is not so tight \citep{izotov_lyman_2021}.

\item The presence of narrow \mgiiL2796,2803 emission close to systemic velocity. \citep{chisholm_optically_2020, xu_tracing_2022}
\end{itemize}

Several recent works propose the use of multi-variable tracers as LyC leakage predictors. Interestingly, large departures from single-variable estimates have been reported \citep[e.g., Fig. 6 of][where several multi-variate models are compared with predictions from \citealp{chisholm_far-ultraviolet_2022}]{jaskot-24b}. Among these models, we highlight:
\begin{itemize}
    \item \cite{mascia_closing_2023} studied a sample of high-$z$ galaxies from the GLASS survey \citep{treu_glass-jwst_2022}, and used a sample of low redshift analogs \citep[from the from the Low-redshift Lyman Continuum Surve, LzLCs][]{flury_low-redshift_2022} to calibrate an empirical relation between observed properties and $\rm f_{esc}^{LyC}$. The relation takes the shape: $\rm log10(f_{esc}^{LyC}) = A + B log_{10}(O32) + C R_{eff}^{UV} + D \beta$, thus their model relies on three parameters (previously identified as individual tracers: $\rm log_{10}(O32)$, $\rm \beta$ and $\rm R_{eff}^{UV}$) to identify LyC leakers.
    \item \cite{lin_empirical_2024} used low-$z$ analogs (LzLCs) to construct an indicator LyC leakage based on the $\rm \beta$ slope and O32 ratios, which is in good agreement with \cite{mascia_closing_2023} predictions. 
    \item \cite{jaskot-24a,jaskot-24b}: use the LzLCs sample to build several predictors (using Cox proportional hazards models) with a broad range of observed parameters, then apply them to $z\sim 3$ and $z\sim 6$ samples. Models including dust attenuation (E(B-V) or $\beta$) and the O32 ratio or morphological measurements ($\rm R_{eff}^{UV}$, $\Sigma_{SFR}$) seem to be the best-performing to predict the $\rm f_{esc}^{LyC}$ for the $z\sim 3$ sample.
\end{itemize}

\newpage

\begin{figure}[t]
\centering
\includegraphics[width=1\columnwidth]{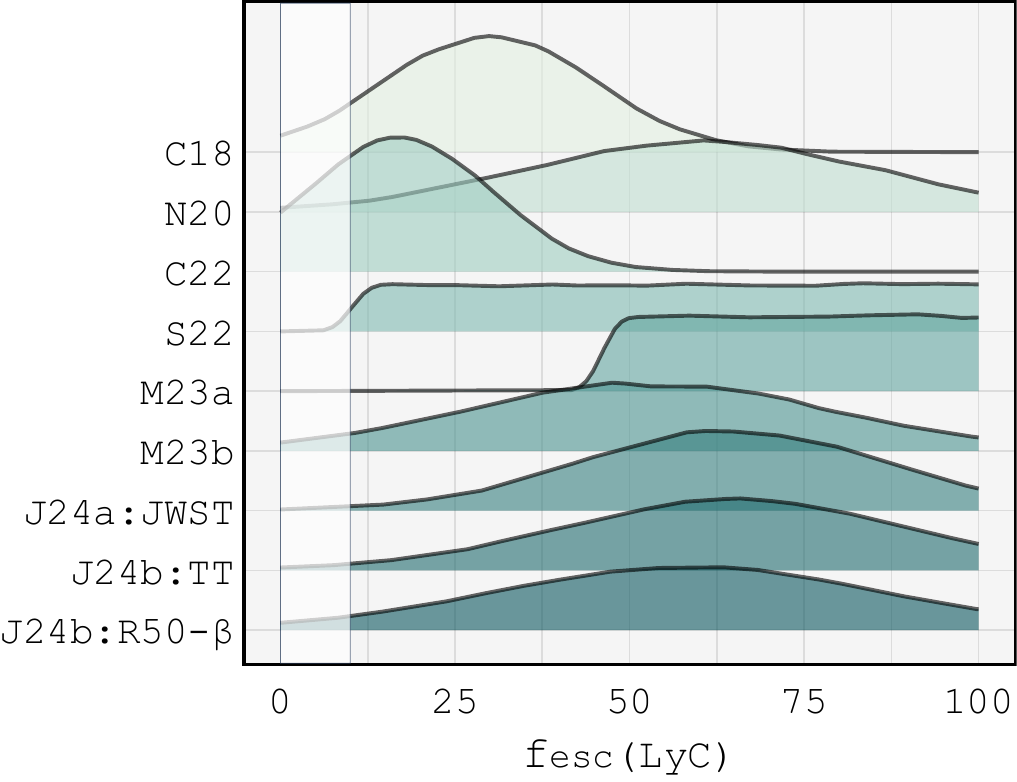}
\caption{\label{fig:fesc_compar} Visualization of the $\rm f_{esc}^{LyC}$ for \gal~predicted by all indicators. We abbreviate the references as follows: C18 for \citet{chisholm_accurately_2018}, N20 for \citet{naidu_rapid_2020}, C22 for \citet{chisholm_far-ultraviolet_2022}, S22 \citet{schaerer_strong_2022}, M23a for \citet{mascia_closing_2023}, M23b for \citet{mascia_new_2023}, J24a:JWST for \citet{jaskot-24b}: `Full JWST', J24b:TT for \citet{jaskot-24a}: `Top Three', J24b:R50-$\beta$ for \citet{jaskot-24a}: `R50-$\beta$'. We show in a white shade the area between $\rm f_{esc}^{LyC} \sim 0 - 10$, as this is where typically non-leakers are located. The median values and corresponding uncertainties are reported in Table \ref{tab:fesc_lyc}.}
\end{figure}

\begin{deluxetable*}{l c r}[ht]
\tabcolsep=3mm
\startdata
\tablecaption{ The estimated $\rm f_{esc}^{LyC}$ of \gal~derived from different indirect indicators. In the case of \cite{jaskot-24a,jaskot-24b} we report the $\rm f_{esc}^{LyC}$ that satisfies $\rm P(f_{esc}^{LyC})<p$ for $\rm p=(0.16,0.50,0.84)$. \label{tab:fesc_lyc}}\\
Reference & Method & $\rm f_{esc}$ \\
\hline
\cite{chisholm_far-ultraviolet_2022} & $\beta$ & $14 \pm 14 \ \%$ \\
\cite{naidu_rapid_2020}              & $\Sigma_{SFR}$ & $60 \pm 25 \ \%$ \\
\cite{chisholm_accurately_2018} & $\rm f_{esc}^{Ly\alpha}$ & $\sim 25-40\ \%$ \\
\cite{schaerer_strong_2022}          & C4C3$^a$ & $> 10 \ \%$ \\
\hline
\cite{mascia_closing_2023}           & $\rm O32$, $\rm \log_{10} R_{eff}^{UV}$, $\beta$ & $> 46 \ \%$ \\
\cite{mascia_new_2023}           & $\rm EW_{0}^{H\beta}$, $\rm \log_{10} R_{eff}^{UV}$, $\beta$ & $53 \pm 26 \ \%$ \\
\cite{jaskot-24a} `Top Three'        & E(B-V), $\rm \log_{10}O32$, $\rm \log_{10}\Sigma_{SFR}$ & $40\% - 60\% - 100\%$ \\
\cite{jaskot-24a} `R50-$\beta$'            & $\beta$, $M_{UV}$, $M^\star$, $\rm \log{10} R_{eff}^{UV}$ & $18\% - 58\% - 100\%$ \\
\cite{jaskot-24b} `Full JWST'        & $M^\star$, $M_{UV}$, $\rm \log_{10}EW^{H\beta}_0$, E(B-V), $\rm 12+\log_{10}O/H$, & $44\% - 65\% - 100 \%$ \\
       & $\rm O32$ , $\rm \log_{10}\Sigma_{SFR}$, $\beta$  & \\
\hline
\cite{lin_empirical_2024}              & $\beta$,O32,$\rm M_{UV}$ & leaker ($\rm P\sim 99\%$) \\
\hline
\enddata
\tablenotetext{a}{\civL1550/\ciiiL1909$>0.75$}
\end{deluxetable*}

\subsection{Estimating the LyC leakage for \gal} \label{sec:fesc}

We now apply the previously defined criteria to \gal. Notably, all single variable from the above requirements are satisfied: a narrow \lya \, which is close to ($\rm \Delta v_{Ly\alpha} = 132 \pm 72 \, km/s$) and has residual emission at the systemic velocity, low dust content ($\rm A_V < 0.1$) and blue $\beta$ slope ($\beta = -2.48\pm 0.23$), is compact and has an elevated value of $\rm \Sigma_{SFR} = 88 \pm 32 \, M_\odot /yr/kpc^2$, presence of bright \civ \, and high \civ/\ciii$=3.3\pm 1.4$ ratio, an elevated O32$>30$ (upper limit, as \oiiL3727,9 is not detected) and the presence of narrow \mgii \, close to the systemic velocity. We report the results for all indicators in Table \ref{tab:fesc_lyc}.

By applying the relations described in \cite{chisholm_accurately_2018,chisholm_far-ultraviolet_2022, naidu_rapid_2020, schaerer_strong_2022}, we recover an escape fraction compatible with the leakage of LyC photons.

We now explore the multi-variate indicators. The relation by \cite{mascia_closing_2023} leads to an elevated lower limit of $\rm f_{esc}^{LyC}$, due to the upper limit in the O32 ratio. A similar answer is found when using \cite{lin_empirical_2024}. This is shown in Fig. \ref{fig:lin_fesc}, where \gal~is located well above the discontinuous line of $75 \%$ probability of LyC leakage (this value is a lower limit, for the reasons above). Using their model, we recover a probability of $\sim 99 \%$ of \gal~being a LyC leaker. Interestingly, the average values for LAEs/SFGs in JADES \citep{kumari_jades_2024} lie along/below the $50 \%$ probability line, respectively. This indicates that LyC leakers are uncommon in their sample, while \gal~shows several qualities in this direction.

By exploring several of the best-performing models from \cite{jaskot-24a,jaskot-24b}, namely: \textit{Top Three}, \textit{R50-$\beta$} and \textit{Full \jwst} models, we derive consistent multi-variable confirmation of the possible LyC nature of \gal. Most observed properties, besides the ones concerning the \lya \, emission, were used in this series of models. All models coincide in assigning a low probability to $\rm f_{esc}^{LyC}<20\%$, and predict mean values of $\rm f_{esc}^{LyC} \sim 60 \%$.

All this information is encoded in Fig. \ref{fig:fesc_compar}, which essentially displays the results of all indicators (Tabel \ref{tab:fesc_lyc}). First of all, is clear that all indicators show evidence of an elevated LyC escape fraction ($\rm f_{esc}^{LyC}>10 \%$). Interestingly, multi-variable indicators (shown downwards in the figure) tend to predict consistently higher $\rm f_{esc}^{LyC}$ when compared to single-variable indicators. Similar effects have been also reported in the literature \citep[e.g.][]{lin_empirical_2024,jaskot-24b}.

Finally, we explored the predictions from the (minimal) model by \cite{maji_predicting_2022}, calibrated using the Sphinx simulation post-processed using RASCAS \citep{michel-dansac_rascas_2020}. However, in the case of \gal, the model over-predicts the escape luminosity. This could be due to \gal~falling out of the SPHINX parameter space (e.g., $\rm SFR_{10}$ and $\rm F^{Ly\alpha}_{esc}$). 

If one message has to be conveyed from these analyses is that estimating $\rm f_{esc}^{LyC}$ with indirect methods is not straightforward. Moreover, calibrations using $\rm f_{esc}^{Ly\alpha}$ could be further affected by the complex physics ruing the properties of the \lya \, line \citep{choustikov_inferring_2024}. Nonetheless, all calibrations and observed properties agree on \gal~having a $\rm f_{esc} \geq 20 \, \%$, thus making it a strong Lyman continuum leaker candidate.

\subsection{Hydrogen ionizing photon production efficiency} \label{sec:xion}
To make $\rm \xi_{ion}$ independent from the estimation of $\rm f_{esc}$, which is indirect and uncertain for the vast majority of high-$z$ galaxies,  we introduce the quantity $\rm \xi_{ion}^0 = \xi_{ion}(f_{esc}^{LyC}=0)$ \citep{lin_empirical_2024}.

In the case of \gal, we report $\rm \log \xi_{ion}^0 = 25.78\pm0.15$ and $\rm \xi_{ion} > 25.84$ assuming $\rm f_{esc} \sim 14 \, \%$ (the one recovered from the $\beta$ slope). This value of $\rm \xi_{ion}$ is elevated with respect to the sample of HAEs studied by \cite{rinaldi_midis_2023} ($\rm \xi_{ion}^0 = 25.5\pm0.1$). In turn, this one was already enhanced relative to the typical $\rm \xi_{ion}$ reported in the literature at lower redshifts \citep[e.g.,][]{lam_mean_2019}. Our high $\rm \xi_{ion}^0$ value inferred for \gal~is however similar to that derived for the $z\sim9$ source studied by \citet{alvarez-marquez_spatially_2024}.

By following the discussion in \citet{rinaldi_midis_2024} and \citet{atek_most_2024}, the average product $\rm f_{esc}\times\xi_{ion}$ can be constrained from the progress of Cosmic Reionization. In the case of \gal, this would impose an $\rm f_{esc} \sim 5 \%$ for $\rm \xi_{ion}^0 = 25.78\pm0.15$, in agreement with what presented by \citet{atek_most_2024}. In the light of our derivations of $\rm f_{esc}^{LyC} >> 5 \%$, we argue that \gal~is {\em caught at the moment of being an efficient reionizer during the EoR}.

\section{Summary and Discussion} \label{sec:conclusion}

We conducted a detailed analysis of the spectroscopic and photometric properties of \gal,  a prominent Ly$\alpha$ emitter at $z=8.279$ selected from the JADES GOODS-N program.  Some of these properties were studied before by \citet{witstok_jades_2024}, with their derived parameter values being in broad agreement with ours in most cases. One notable exception is the UV $\beta$ slope, though, for which we obtained a lower value ($\beta=-2.2 - -2.7$) than  \citet{witstok_jades_2024}, albeit consistent within the error bars. The reason for this departure might stem from the fact that we used the best-fit FUV continuum to determine $\beta$, while \citet{witstok_jades_2024} made use of the masked spectrum. Remarkably, \gal~is a low stellar-mass galaxy ($M^\ast \sim 10^{7.66} \, \rm M_\odot$), contrary to the vast majority of Ly$\alpha$ emitters known at $z>7$, which are typically significantly more luminous and massive.

Although \gal~is at too high a redshift to detect a significant amount of ionizing photons, we can conclude that this source is a robust candidate for a LyC leaker, namely an early reionizer. Indeed, our analysis of multiple indirect tracers univocally indicates that the $\rm f_{esc}$ of \gal~must be $>10\%$, i.e. well above the average threshold needed for galaxies to drive the reionization process \citep{atek_most_2024}. Particularly, \gal~is a strong CIV] emitter, as expected for LyC leakers \citep{schaerer_strong_2022}.

We also reported the detection of the NIII]$\lambda 1749-1753$ emission line complex in \gal, which was not identified in previous studies of this source. Very recently, NIII]$\lambda 1749-1753$ was discovered in a handful of high-$z$ galaxies \citep[e.g.,][]{marques-chaves_extreme_2024, schaerer_discovery_2024}, but our source is the first reported case of simultaneous NIII] and \lya \, emission at the EoR. As in \citet{schaerer_discovery_2024}, our measured NIII]$\lambda 1749-1753$ flux implies a supra-solar N abundance, which is at odds with the sub-solar metallicity derived for our target. 

The exact origin of the NIII]$\lambda 1749-1753$ is unclear, but \citet{schaerer_discovery_2024} suggested that there could be a link between this emission and extremely high values of $\Sigma_{\rm M^\star}$ and $\Sigma_{\rm SFR}$. This is confirmed for our source, which has $\rm \Sigma_{M^\star} \sim 355 \, \rm M_\odot / pc^2$ and $\Sigma_{\rm SFR} \sim 88 \, \rm M_\odot /yr/kpc^2$ and, thus, is placed on the $\Sigma_{\rm SFR}$-$\Sigma_{\rm M^\star}$ surface density relation derived by \citet{morishita_enhanced_2024}, along with other extreme density sources like previously found N-emitters. The  NIII]$\lambda 1749-1753$ could also be associated with the presence of Wolf-Rayet stars injecting N to the ISM \citep{rivera-thorsen_sunburst_2024}, but in this case we would also expect to see the permitted NIII$\lambda\lambda 4620, 4640$ lines and broadened HeII$\lambda 4686$ emission, which we do not detect in \gal.

In summary, \gal~is, to our knowledge, the most robust candidate found to date for a source being caught at the moment of reionizing its surrounding medium. {\em Which physical processes could lead to the properties observed in \gal?}  Based on magnetohydrodynamical galaxy simulations, \citet{witten_deciphering_2024}  concluded that prominent Ly$\alpha$ emission at the EoR could be explained via galaxy mergers inducing vigorous star formation. This remains to be investigated from the observational point of view.  Further \jwst \,  observations should allow us to find more sources like \gal~in the early Universe and thoroughly study their properties, eventually leading to the unlock of the process of reionization.

\section{ACKNOWLEDGMENTS} \label{sec:ACKNOWLEDGMENTS}

We thank Pratika Dayal for their useful discussions. RN and KIC acknowledge funding from the Dutch Research Council (NWO) through the award of the Vici Grant VI.C.212.036. E.I. and P.R. acknowledge funding from the Netherlands Research School for Astronomy (NOVA). This work is based on observations made with the NASA/ESA/CSA James Webb Space Telescope. The data were obtained from the Mikulski Archive for Space Telescopes at the Space Telescope Science Institute, which is operated by the Association of Universities for Research in Astronomy, Inc., under NASA contract NAS 5-03127 for \textit{JWST}. These observations are associated with \textit{JWST} programs GTO \#1180, GO \#1210, GO \#1963 and GO \#1895. The authors acknowledge the team led by coPIs D. Eisenstein and N. Luetzgendorf for developing their respective observing programs with a zero-exclusive-access period. Also based on observations made with the NASA/ESA Hubble Space Telescope obtained from the Space Telescope Science Institute, which is operated by the Association of Universities for Research in Astronomy, Inc., under NASA contract NAS 526555.

\textit{Software}: \texttt{AstroPy} \citep{collaboration_astropy_2022}, \texttt{BAGPIPES} \citep{carnall_how_2019}, \texttt{dustmaps} \citep{green_dustmaps_2018}, \texttt{Matplotlib} \citep{hunter_matplotlib_2007}, \texttt{NumPy} \citep{harris_array_2020}, \texttt{Photutils} \citep{bradley_astropyphotutils_2022}, \textsc{Python} \citep{van_rossum_python_1995}, \texttt{SciPy} \citep{virtanen_scipy_2020}, \texttt{TOPCAT} \citep{taylor_topcat_2017}, \texttt{WebbPSF} \citep{perrin_updated_2014}.

\textit{Facilities}: \textit{HST}, \textit{JWST}


\bibliography{references}{}
\bibliographystyle{aasjournal}

\end{document}